\documentclass[prd,twocolumn,showpacs,showkeys,preprintnumbers,amsmath,amssymb,floatfix]{revtex4}

\usepackage[utf8]{inputenc}
\usepackage{hyperref}

\usepackage{dcolumn}
\usepackage{bm}
\usepackage{graphicx}
\usepackage{subfigure}

\def\be{\begin{equation}}
\def\ee{\end{equation}}
\def\bestar{\begin{equation*}}
\def\eestar{\end{equation*}}

\def\p{\partial}

\begin{document}

\title{Thermodynamics of a higher dimensional noncommutative anti-de Sitter-Einstein-Born-Infeld black hole}

\author{Ang\'elica Gonz\'alez}
\email{gombstar@xanum.uam.mx}
\affiliation{Departamento de F\'{i}sica \\ Universidad Aut\'{o}noma Metropolitana - Iztapalapa\\
                   Av. San Rafael Atlixco 186, A.P. 55-534, C.P. 09340, Ciudad de México, M\'exico}

\author{Rom\'an Linares}
\email{lirr@xanum.uam.mx}
\affiliation{Departamento de F\'{i}sica \\ Universidad Aut\'{o}noma Metropolitana - Iztapalapa\\
                  Av. San Rafael Atlixco 186, A.P. 55-534, C.P. 09340, Ciudad de México, M\'exico}

\author{Marco Maceda}
\email{mmac@xanum.uam.mx}
\affiliation{Departamento de F\'{i}sica \\ Universidad Aut\'{o}noma Metropolitana - Iztapalapa\\
                   Av. San Rafael Atlixco 186, A.P. 55-534, C.P. 09340, Ciudad de México, M\'exico}
                   
\author{Oscar S\'anchez-Santos}
\email{oscarss@xanum.uam.mx}
\affiliation{Departamento de F\'{i}sica \\ Universidad Aut\'{o}noma Metropolitana - Iztapalapa\\
                   Av. San Rafael Atlixco 186, A.P. 55-534, C.P. 09340, Ciudad de México, M\'exico}

\date{\today}

\begin{abstract}
We analyze noncommutative deformations of a higher dimensional anti-de Sitter-Einstein-Born-Infeld black hole. Two models based on noncommutative inspired distributions of mass and charge are discussed and their thermodynamical properties such as the equation of state are explicitly calculated. In the (3+1)-dimensional case the Gibbs energy function of each model is used to discuss the presence of phase transitions.
\end{abstract}


\maketitle

\section{Introduction}
\label{intro}

In noncommutative quantum field theory, the use of coherent states has allowed the implementation of commutation relations among coordinates without violating Lorentz invariance~\cite{Smailagic:2003rp,Smailagic:2004yy}; further developments have established a deep relationship with the noncommutative Voros product~\cite{Banerjee:2009xx}. The main lesson learned from these works is that point-like sources should be replaced by smeared distributions, actually Gaussian distributions, depending on the noncommutative parameter $\theta$; in the limit of vanishing $\theta$, the standard expressions for Green's functions in quantum field theory are recovered.
 
These ideas have been applied recently to gravitational systems~\cite{Nicolini:2005vd,Nicolini:2008aj,Tejeiro:2010gu,Liang:2012vx,Rahaman:2013gw,Gonzalez:2014mza} and the analysis of classical black holes, such as the Schwarzschild or the Reissner-Nordström (RN) spacetime, has been performed. As it is well known, these two black holes are special cases of more general models involving nonlinear electrodynamics that reduce to Maxwell's electrodynamics in certain limits. 

Among them, Born-Infeld (BI) electrodynamics~\cite{Born:1934ji,Born:1934gh} has received a lot of attention since its formulation, arising in several contexts such as in the analysis of the low-energy effective action for an open superstring~\cite{Fradkin:1985qd} or in the low-energy dynamics of D-branes~\cite{Leigh:1989jq}; different aspects of the associated black holes have been studied in the literature~\cite{Wiltshire:1988uq,Rasheed:1997ns,Tseytlin:1999dj,Tamaki:2000ec,Gibbons:2001gy,Breton:2003tk,Aiello:2004rz,Cataldo:1999wr,Fernando:2003tz,Cai:2004eh,Dey:2004yt}.
Additionally, in the presence of a gravitational field, the Einstein field equations admit a black hole solution~\cite{GarciaD2007};  in the static case, the EBIon is the corresponding spherically symmetric solution~\cite{Demianski:1986wx}. These black hole solutions are the nonlinear generalization of the RN black hole solutions; they are described by the mass $M$ and the charge $Q$ of the black hole in addition to the BI parameter $b$, directly related to the strength of the electromagnetic field at the position of the charge. 

In this paper we construct two noncommutative models of anti-de Sitter-Einstein-Born-Infeld (adSEBI) spacetime using smeared distributions for the mass and charge of the black hole; the sources of the gravitational field have then a noncommutative origin. The interest on these models lies first on the natural generalization of well-known results on black hole thermodynamics using Maxwell's electrodynamics. Additionally, this approach allows us to introduce a fundamental length scale that can be related to quantum gravity effects and black holes should certainly be affected by them~\cite{Dvali:2010gv,Spallucci:2014kua}.  

The models provide also a nice illustration of the changing behavior of thermodynamical quantities when the noncommutative parameter is turn on; the influence on critical points is clearly seen and phase transitions also exist. Moreover, explicit calculations can be performed in spite of the nontrivial nature of both the nonlinear electrodynamics and the noncommutative deformation; the amount of work involved in obtaining a solution to the noncommutative gravitational field equations is similar to that of the usual case. This represents an important advantage for the study of quantum effects on gravity since we have then a nonpertubative treatment of black holes in general that allows close comparison with the standard results.

The paper is organised as follows: In Sec.~\ref{sec2}, we review the Einstein-Born-Infeld (EBI) solution and formulate two noncommutative models, first by considering a noncommutative distribution of mass and then by using a noncommutative distribution of charge; the analysis of the metric is presented for both cases. The thermodynamical properties of each of these models are then analyzed in Sec.~\ref{sec3}; the equation of state and Gibbs energy function for the (3+1)-dimensional case are also found. We end up with some remarks and future work in the Conclusions. Throughout this paper we will use units where $G = c = 1$.

\section{Noncommutative modifications of the field equations}
\label{sec2}

The noncommutative field equations for the model can be written down straightforwardly. First, we recall that the action of a nonlinear electrodynamics coupled to gravity with cosmological constant $\Lambda$ in $(d+1)$-dimensions is given by
\be
S = \frac 1{16\pi} \int \, d^{d+1}x \sqrt{-g} \left[ R -2\Lambda + L(F) \right],
\ee
where $R$ is the scalar curvature calculated from the metric $g_{\mu\nu}$, $g := det (-g_{\mu\nu})$ and $L(F)$ is an arbitrary function of the electromagnetic tensor $F_{\mu\nu}$. Since Maxwell's electrodynamics should be valid in the weak field limit we must have
\be
L(F) = - F^{\mu\nu} F_{\mu\nu} + O(F^4),
\ee
to lowest order. In the specific case of BI electrodynamics, the function $L(F)$ has the form
\be
L(F) = 4b^2 \left( 1 - \sqrt{1+ \frac {F^{\mu\nu} F_{\mu\nu}}{2b^2}} \right),
\ee
where $b$ is the BI parameter. The classical EBI field and conservation equations derived from the above action are then~\cite{Fernando:2003tz}
\begin{eqnarray}
&&\nabla_\mu \left( \frac {F^{\mu\nu}}{\sqrt{1 + F^2/2b^2}} \right) = 0,
\nonumber \\[4pt]
&&R_{\mu\nu} = \Lambda g_{\mu\nu} + \frac {\partial L(F)}{\partial g^{\mu\nu}} 
\nonumber \\[4pt]
&&
+ \frac 12 g_{\mu\nu} \left[ - L(F) + \frac {\partial L(F)}{\partial g^{\rho\sigma}} g^{\rho\sigma} \right],
\label{classfequ}
\end{eqnarray}
where
\be
 \frac {\partial L(F)}{\partial g^{\mu\nu}} = - \frac {2 F_{\sigma\nu} F^\sigma{}_\mu}{\sqrt{1+ F^2/2b^2}}.
\ee

Let us consider a static spherically symmetric spacetime in $(d+1)$-dimensions
\be
ds^2 = - e^{2\mu(r)} dt^2 + e^{2\nu(r)} dr^2 + r^2 ds_{(d-1)}^2,
\ee
where 
\begin{eqnarray}
ds_{(d-1)}^2 &=& d\theta_1^2 + \sin^2 \theta_1 d\theta_2^2 + \dots + \sin^2 \theta_1 \cdots \sin^2 \theta_{d-2} d\theta_{d-1}^2
\nonumber \\[4pt]
&=& \sum_{j=1}^{d-1} \left( \prod_{i=1}^{j-1} \sin^2 \theta_i \right) d\theta_j^2,
\end{eqnarray}
is the line element on the $S^{d-1}$ sphere with total area (or $(d-1)$-dimensional solid angle)
\be
\Omega_{(d-1)} = \frac {2\pi^{d/2}}{\Gamma(d/2)}.
\ee
Using the orthonormal basis
\be
\theta^0 = e^{\mu(r)} dt, \quad \theta^1 = e^{\nu(r)} dr, \quad \theta^{a+1} = r \left( \prod_{i=1}^{a-1} \sin \theta_i \right) d\theta_a, 
\ee
with $a=1, \dots, d-1$, the Ricci components are found to be
\begin{eqnarray}
R_{00} &=& e^{-2\nu} \left( \mu^{\prime\prime} + \mu^{\prime\,2} + \mu^\prime\nu^\prime + \frac {d-1}r \mu^\prime \right),
\nonumber \\[4pt]
R_{11} &=& e^{-2\nu} \left( -\mu^{\prime\prime} - \mu^{\prime\,2} + \mu^\prime\nu^\prime + \frac {d-1}r \nu^\prime \right),
\nonumber \\[4pt]
R_{a+1a+1} &=& \frac {d-2}{r^2} (1 - e^{-2\nu}) + \frac 1r (-\mu^\prime + \nu^\prime) e^{-2\nu}.
\end{eqnarray}
We now formulate two noncommutative models.

\subsection{Model A: Noncommutative mass deformation}
\label{subsec1}

At this level, noncommutativity is introduced in the model by defining a noncommutative energy-matter tensor 
\be
\mathfrak T^\mu{}_{\nu} := diag (h_1, h_1, h_3, \dots, h_3), \quad h_3 := (r^2 h_1)_{,r}/2r, 
\ee
where
\be
h_1 := -\rho_m(r) = -\frac M{(4\pi\theta)^{d/2}} e^{-r^2/4\theta},
\ee
and a comma denotes partial derivative with respect to the radial coordinate $r$; the normalization of $\rho_m(r)$ is such that
\be
\int d^d x \, \rho_m (r) = M.
\ee
It follows that
\be
\mathfrak T_{\mu\nu} = diag( -h_1, h_1, h_3, \dots , h_3),
\ee
in the orthonormal basis given previously. The function $\rho_m$ in the previous expressions represents a non-commutative diffused distribution of matter which in the commutative limit $\theta \to 0$ becomes a point-like source of mass $M$ located at the origin. It can be verified straightforwardly that this form of the energy-momentum tensor satisfies $\nabla_\mu \mathfrak T^{\mu\nu} = 0$ due to the definition of the function $h_3$ in terms of $h_1$.

The field equations we are interested in solving are 
\begin{eqnarray}
R_{\mu\nu} -\Lambda g_{\mu\nu} &=& \frac {\partial L(F)}{\partial g^{\mu\nu}} + \frac 12 g_{\mu\nu} \left( - L(F) + \frac {\partial L(F)}{\partial g^{\rho\sigma}} g^{\rho\sigma} \right) 
\nonumber \\[4pt]
&&+ 8\pi \left( \mathfrak T_{\mu\nu} - \frac 12 g_{\mu\nu} \mathfrak T^\lambda{}_{\lambda} \right),
\end{eqnarray}
for the purely electric case where $F_{tr} = E(r)$ in the coordinate basis. At this stage the conservation laws in Eqs.~(\ref{classfequ}) are not being changed from their classical expressions; this will be done in Sec.~\ref{subsec2} by means of a noncommutative electric current.

Let us proceed to the solution for the static electric charged case. From the BI Lagrangian we have~\cite{Fernando:2003tz}
\be
\frac {\partial L(F)}{\partial g^{00}} = -\frac {\partial L(F)}{\partial g^{11}},\qquad \frac {\partial L(F)}{\partial g^{22}} = \frac {\partial L(F)}{\partial g^{33}} = 0.
\label{der00der11bi}
\ee
These expressions are a direct consequence of the form of the BI Lagrangian. From them it follows
\be
R_{00} = - R_{11} \quad \Rightarrow \quad \mu + \nu = 0.
\label{r00r11rel}
\ee
In consequence, we only need to determine one metric function. From the field equations for the $aa$-components we obtain 
\be 
\frac {d-2}{r^2}  - \frac 1{r^{d-1}} (r^{d-2} e^{-2\nu})_{,r} = \Lambda -\frac 12 L(F) + \frac {\partial L(F)}{\partial g^{00}} g^{00} - 8\pi h_1.
\ee
Since the conservation law is the same as in the classical case, we have
\be
L(F) = 4b^2 \left( 1 - \sqrt{1 - E(r)^2/b^2} \right),
\ee
where 
\be
E(r) = \frac Q{\sqrt{r^{2(d-1)} + Q^2/b^2}},
\ee
is the electric field in the coordinate basis due to a charge $Q$ located at the origin. Therefore, we arrive to the following differential equation for the function $\nu$
\begin{widetext}
\be
\frac {d-2}{r^2}- \frac {(r^{d - 2} e^{-2\nu})_{,r}}{r^{d-1}} = \Lambda - \frac 2{r^{d-1}} b \left( b r^{d-1}  - \sqrt{Q^2 + b^2 r^{2(d-1)}} \right) + \frac {8\pi M}{(4\pi\theta)^{d/2}} e^{-r^2/4\theta},          
\ee
or equivalently
\be
(r^{d-2} e^{-2\nu})_{,r} = (d-2) r^{d-3} - \Lambda r^{d-1} + 2 b ( b r^{d-1} - \sqrt{Q^2 + b^2 r^{2(d-1)}} ) -\frac {8\pi M}{(4\pi\theta)^{d/2}}  r^{d-1} e^{-r^2/4\theta}.
\label{diffequmodela}
\ee
Straightforward integration gives then
\begin{eqnarray}
e^{-2\nu} &=& 1 + \frac {b_0}{r^{d-2}} - \frac 1d \Lambda r^2 + \frac 2d b^2 r^2 \left( 1- \sqrt{1 + \frac {Q^2}{b^2 r^{2(d-1)}}} \right) + \frac {2 (d-1)}d \frac {Q^2}{r^{d-2}}  
\nonumber \\[4pt]
&&\times \int_r^\infty \frac {ds}{\sqrt{s^{2(d-1)} + Q^2/b^2}} - \frac {8\pi M}{2\pi^{d/2}} \frac 1{r^{d-2}} \gamma\left( \frac d2, \frac {r^2}{4\theta} \right),
\label{ncsolelectric}
\end{eqnarray}
where $\gamma(n,z)$ is the lower gamma function
\be
\gamma(n,z) := \int_0^z dt \, t^{n-1} e^{-t},
\ee
and $b_0$ is an arbitrary constant. Using the identity~\cite{Gunasekaran:2012dq}
\be
{}_2 F_1 \left( \frac 12, c ; c+1; - z \right) = c \int_0^1 \frac {t^{c-1} dt}{(1 + zt)^{1/2}},
\ee
where $c$ is a constant, the equivalent expression
\begin{eqnarray}
e^{-2\nu} &=& 1 + \frac {b_0}{r^{d-2}} - \frac 1d \Lambda r^2 + \frac {2b^2 r^2}d \left(1 - \sqrt{1 + \frac {Q^2}{b^2 r^{2(d-1)}}} \right) + \frac {2(d-1)}{d(d-2)} \frac {Q^2}{r^{2(d-2)}}
\nonumber \\[4pt]
&& \times {}_2F_1\left( \frac 12, \frac {d-2}{2(d-1)}; \frac {3d-4}{2(d-1)}; - \frac {Q^2}{b^2 r^{2(d-1)}}\right) 
- \frac {8\pi M}{2\pi^{d/2}} \frac 1{r^{d-2}} \gamma\left( \frac d2, \frac {r^2}{4\theta} \right),
\label{lapsea}
\end{eqnarray}
\end{widetext}
for the solution is obtained. Usually $b_0 = -2M$, where $M$ is the mass of the black hole; in the following we set $b_0 = 0$ and therefore, a black hole regular at the origin exists. 

\subsection{Model B: Noncommutative BI electrodynamics}
\label{subsec2}

We now proceed to analyze the noncommutative deformation of the BI solution for the electromagnetic field. For this purpose, the conservation law in Eq.~(\ref{classfequ}) is modified to
\be
\nabla_\mu \left( \frac {F^{\mu\nu}}{\sqrt{1 + F^2/2b^2}} \right) = \mathfrak J^\nu,
\ee
where
\be
\mathfrak J^\nu = [ \rho_Q (r), \vec 0] :=  \left[ \frac Q{(4\pi\theta)^{d/2}} e^{-r^2/4\theta}, \vec 0 \right],
\ee
is the noncommutative 4-vector current; the point-like charge has been replaced by a smooth distribution involving the noncommutative parameter $\theta$.

With the Ansatz $F_{tr} = E(r)$, the only nontrivial conservation law to be considered is
\be
\partial_r \left( \frac {e^{-2(\mu + \nu)} r^{d-1} E(r)}{\sqrt{1 - E(r)^2/b^2}} \right) = -\frac Q{(4\pi\theta)^{d/2}} e^{\mu+\nu} r^{d-1} e^{-r^2/4\theta}.
\ee
We use now the fact that Eq.~(\ref{r00r11rel}) is still valid for the noncommutative model since we are only introducing a charge density without modifying the form of the BI Lagrangian. In consequence $\mu + \nu = 0$ and thus the previous equation becomes
\be
\partial_r \left( \frac {r^{d-1} E(r)}{\sqrt{1 - E(r)^2/b^2}} \right) = -\frac Q{(4\pi\theta)^{d/2}} r^{d-1} e^{-r^2/4\theta}.
\ee
It follows that
\be
E (r) = \frac {H(r)}{\sqrt{r^{2(d-1)} + H(r)^2/b^2}},
\ee
where 
\begin{eqnarray}
H(r) &:=& \frac Q{(4\pi\theta)^{d/2}} \int_0^r dz \, z^{d-1} e^{-z^2/4\theta} 
\nonumber \\[4pt]
&=& \frac Q{2\pi^{d/2}} \gamma\left( \frac d2, \frac {r^2}{4\theta} \right).
\end{eqnarray}
When coupled with the gravitational field, and without  performing a noncommutative deformation of the mass, we obtain the following differential equation for the metric function $\nu$
\begin{widetext}
\begin{eqnarray}
(r^{d-2} e^{-2\nu})_{,r} &=& (d - 2) r^{d-3} - \Lambda r^{d-1} + 2b^2 \left( r^{d-1} - \sqrt{\frac {H^2(r)}{b^2} + r^{2(d-1)}} \right).
\label{diffequmodelb}
\end{eqnarray}
The general solution can then be written as
\begin{eqnarray}
e^{-2\nu} &=& 1 + \frac {b_0}{r^{d-2}} - \frac 1d \Lambda r^2 + 2b^2 \int_r^\infty ds \left( \sqrt{\frac {H^2(s)}{b^2} + s^{2(d-1)}} - s^{d-1} \right),
\end{eqnarray}
or equivalently
\be
e^{-2\nu} = 1 - \frac {2M}{r^{d-2}} - \frac 1d \Lambda r^2 + \frac 2d b^2 r^2 \left( 1- \sqrt{1 + \frac {H^2(r)}{b^2 r^{2(d-1)}} }\right) 
+ \frac 2d \frac 1{r^{d-2}} \int_r^\infty ds \frac { - H(s) s^d \rho_Q(s) + (d -1) H^2(s)}{\sqrt{s^{2(d-1)} + H^2(s)/b^2}},
\label{lapseb}
\ee
\end{widetext}
where we have set $b_0 = -2M$; with this choice the metric coefficient $g_{00}$ will blow up at the origin. It can be seen that the well known expression for EBI spacetime is recovered in the limit $\theta \to 0$.

\section{Thermodynamical properties}
\label{sec3}

For each of the black holes solutions in the previous section, it is straightforward to obtain their thermodynamical properties. The first step is to calculate the energy density surface for $r = R = const.$; it is given by~\cite{Brown:1994gs} 
\be
\epsilon(R) := \frac {k(R)}{8\pi} - \epsilon_0 (R),
\ee
where $\epsilon_0(r)$ is a energy density surface that can be conveniently chosen, and $k$ is the trace of the extrinsic curvature $k_{ab}$ of the $(d-1)$-hypersurface defined by the metric $r^2 ds_{(d-1)}^2$. Furthermore, by writing the metric $ds^2$ for $r=const.$ as 
\be
-N^2 dt^2 + \sigma_{ab} (dx^a + V^a dt)(dx^b + V^b dt), 
\ee
we associate to it some metric functions $\sigma_{ab}$ and a shift vector $V^a$. 

Once the energy density surface $\epsilon(R)$ is known, the corresponding total quasilocal energy can be found from 
\be
E(R) = \int d^{(d-1)} x \, \sqrt{\sigma} \epsilon(R),
\ee 
where the integration is over the $(d-1)$-dimensional hypersurface defined by $r^2 ds_{(d-1)}^2$, and $\sigma := det (\sigma_{ab})$; this is the thermodynamic internal energy within the boundary $R$. Using the above expressions, the temperature~\cite{Brown:1994gs}
\be
T(R) := \left( \frac {\p E}{\p S} \right) = \frac 1{2\pi} \frac {\kappa_H}{N(R)},
\ee
the pressure
\be
{\cal P} := - \left( \frac {\partial E(R)}{\partial (\Omega_{(d-1)} R^{d-1})} \right),
\label{termopressure}
\ee
and the specific heat
\be
C_R := \left( \frac {\partial E}{\partial T} \right) = \left( \frac {\partial E(R)}{\partial r_{H}} \right) \Big/ \left( \frac {\partial T(R)}{\partial r_{H}} \right), 
\ee
can be found; here $\kappa_H$ is the surface gravity at the horizon $r_H$, $S := \Omega_{(d-1)} r_H^{d-1}/4$ is the entropy of the black hole and in the last expression above, it is assumed that the condition
\be
N^2(r_H) =  - e^{2\mu(r_H)} = - e^{-2\nu(r_H)} = 0,
\label{horizonequ}
\ee 
on the lapse function has been used to write the mass $M$ as a function of $r_H$ and $R$, i.e. $M = M(r_H, R)$. In the following, we discuss the thermodynamical properties of the noncommutative models given in the previous section.

\subsection{Model A}
\label{subsec3}

First, we review very briefly the calculation of the total quasilocal energy $E(R)$; we have
\be
\epsilon(R) = - \frac 1{8\pi} \frac {2N(R)}R - \epsilon_0(R) = - \frac 1{4\pi} \frac {N(R)}R - \epsilon_0(R),
\ee
and in consequence
\be
E(R) = - R N(R) - 4\pi R^2\epsilon_0(R).
\ee
The function $\epsilon_0(R)$ is chosen as 
\be
\epsilon_0(R) = -\frac 1{4\pi R} \sqrt{1 - \frac 13 \Lambda R^2},
\label{epsilon0}
\ee
implying that $\epsilon(R)$ vanishes identically for anti-de Sitter spacetime ($M = 0 = Q$). The thermodynamic surface pressure is then according to Eq.~(\ref{termopressure})
\be
{\cal P} = \frac {\p}{\p (4\pi R^2)} (R N(R)) + \frac {\p (R^2 \epsilon_0 (R))}{\p R^2}.
\ee
In this model the lapse function $N^2(r)$ is given by the right hand side of Eq.~(\ref{lapsea}). 
\begin{widetext}
From the condition $N(r_H) = 0$, the mass $M$ is written as 
\begin{eqnarray}
8\pi M &=& \frac {2 \pi^{d/2} r_H^{d-2}}{\gamma \left( \frac d2, \frac {r_H^2}{4\theta} \right)} \left[ 1 - \frac 1d \Lambda r_H^2 + \frac 2d b^2 r_H^2 \left( 1 - \sqrt{1 + \frac {Q^2}{b^2 r_H^{2(d-1)}}} \right) + \frac {2(d-1)}{d(d-2)} \frac {Q^2}{r_H^{2(d-2)}}  \right.
\nonumber \\[4pt]
&&\left. \times {}_2F_1\left( \frac 12, \frac {d-2}{2(d-1)}; \frac {3d-4}{2(d-1)}; - \frac {Q^2}{b^2 r_H^{2(d-1)}}\right) \right].
\end{eqnarray}
On the other hand, from Eq.~(\ref{diffequmodela}) evaluated at the horizon $r_H$, we obtain the surface gravity of the black hole as
\be
\kappa_H = \frac 1{2r_H^{d-2}} \left[ (d-2) r_H^{d-3} - \Lambda r_H^{d-1} + 2b^2 r_H^{d-1} \left( 1 - \sqrt{1 + \frac {Q^2}{b^2 r_H^{2(d-1)}}} \right) - \frac {8\pi M}{(4\pi\theta)^{d/2}} r_H^{d-1} e^{-r_H^2/4\theta} \right].
\label{kappamodela}
\ee
Let us specialize the above expressions to the (3+1)-dimensional case; we have that the temperature at the hypersurface $r = R = const. > r_H$ is (see Appendix~\ref{appa})
\be
T(R) = \frac 1{N(R)} \frac {\displaystyle \gamma \left( \frac 32, \frac {R^2}{4\theta} \right) }{\displaystyle \gamma \left( \frac 32, \frac {r_H^2}{4\theta} \right) } \frac 1{4\pi r_H} \left[ 1 - \Lambda r_H^2 + 2b^2 r_H^2 \left( 1 - \sqrt{1+ \frac {Q^2}{b^2 r_H ^4}} \right) - \frac {8\pi M}{(4\pi\theta)^{3/2}} r_H^2 e^{-r_H^2/4\theta} \right],
\label{temp1modela}
\ee
where
\be
N^2(R) = 1 - \frac 13 \Lambda R^2 + \frac 23 b^2 R^2 \left( 1- \sqrt{1 + \frac {Q^2}{b^2 R^4}} \right) +  \frac {4 Q^2}{3 R^2} {}_2 F_1 \left( \frac 12, \frac 14; \frac 54 ; -\frac {Q^2}{b^2 R^4} \right) - \frac {8\pi M}{2\pi^{3/2}} \frac 1R \gamma \left( \frac 32, \frac {R^2}{4\theta} \right),
\ee 
and 
\be
8\pi M = \frac {2 \pi^{3/2} r_H}{\displaystyle \gamma \left( \frac 32, \frac {r_H^2}{4\theta} \right)} \left[ 1 - \frac 13 \Lambda r_H^2 + \frac 23 b^2 r_H^2 \left( 1- \sqrt{1 + \frac {Q^2}{b^2 r_H^4}} \right) +  \frac {4 Q^2}{3 r_H^2} {}_2 F_1 \left( \frac 12, \frac 14; \frac 54 ; -\frac {Q^2}{b^2 r_H^4} \right) \right].
\label{massatrh}
\ee
\end{widetext}
Notice the presence of the incomplete gamma functions in the expression for the temperature Eq.~(\ref{temp1modela}). This is a nontrivial modification with respect to the standard case and is due to the noncommutative deformation: the ratio of the incomplete gamma functions becomes unity in the commutative limit $\theta \to 0$. 

\subsubsection{Equation of state}

\begin{figure}[pbth]
\centering
\includegraphics[width=0.45\textwidth]{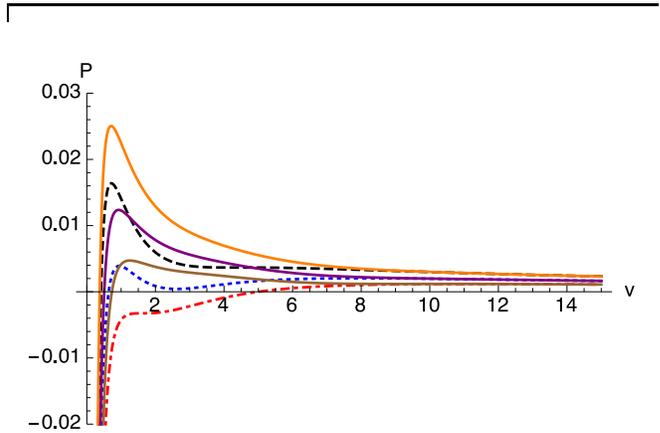}
\caption{$P-v$ diagram of noncommutative BI black hole (mass deformation). The thick lines correspond to the noncommutative equation of state with $\theta = 1$ for $T = 0.04517, 0.035$ and $0.026885$ (top to bottom); the dashed lines are the corresponding commutative cases in the same order. We have set $Q=1, M=10$ and $b = 0.45$.}  
\label{fig1}
\end{figure}

Let us now proceed to find the equation of state for our black hole solution in (3+1)-dimensions; we follow~\cite{Gunasekaran:2012dq}. We know that the entropy of the solution is given by
\be
S = \frac 14 A, \qquad A := 4\pi r_H^2,
\ee
with thermodynamic volume and associated pressure
\be
V := \frac {4\pi}3 r_H^3, \qquad P := -\frac 1{8\pi} \Lambda,
\ee
respectively. Then, by taking $T := \kappa_H/2\pi$, we have
\be
P = \frac Tv - \frac 1{2\pi v^2} - \frac {b^2}{4\pi} \left(1 - \sqrt{1+ \frac {16Q^2}{b^2 v^4}} \right) + \frac M{(4\pi\theta)^{3/2}} e^{-v^2/16\theta},
\ee
where the specific volume $v$ is defined as $v := 2 r_H$. A typical $P-v$ diagram for a fixed mass $M$ is shown in Fig.~\ref{fig1}. Notice the influence of noncommutativity (solid line) on the locus of critical points.

Critical points are defined by the conditions
\be
\left( \frac {\p P}{\p v} \right)_{T_c} = 0, \qquad \left( \frac {\p^2 P}{\p v^2} \right)_{T_c} = 0.
\ee
In the commutative case these two conditions translate into a cubic algebraic equation for the specific volume $v$. In our model it becomes the trascendental equation
\begin{eqnarray}
x^3 + px + q = a e^{-\frac 1{16 \theta}\sqrt{\frac 1{x^2} - \frac {16 Q^2}{b^2}}} \left( \frac 1{x^2} - \frac {16 Q^2}{b^2} \right) 
\nonumber \\[4pt]
\times \left( \sqrt{\frac 1{x^2} - \frac {16 Q^2}{b^2}} - 24 \theta \right),
\label{critpmodela}
\end{eqnarray}
where
\be
p := - \frac {3b^2}{2^5 Q^2}, \quad q := \frac {b^2}{2^8 Q^4}, \quad a:= \frac {8\pi Mb^2}{2^{20} Q^4 \pi^{3/2} \theta^{7/2}}. 
\ee
and 
\be
x^{-2} := v^4 + \frac {16 Q^2}{b^2}, \qquad |x| \leq b/4Q.
\label{xvmodela}
\ee
In the limit $\theta \to 0$, the right hand side of Eq.~(\ref{critpmodela}) vanishes.

\begin{figure}[h]
\centering
\includegraphics[width=0.45\textwidth]{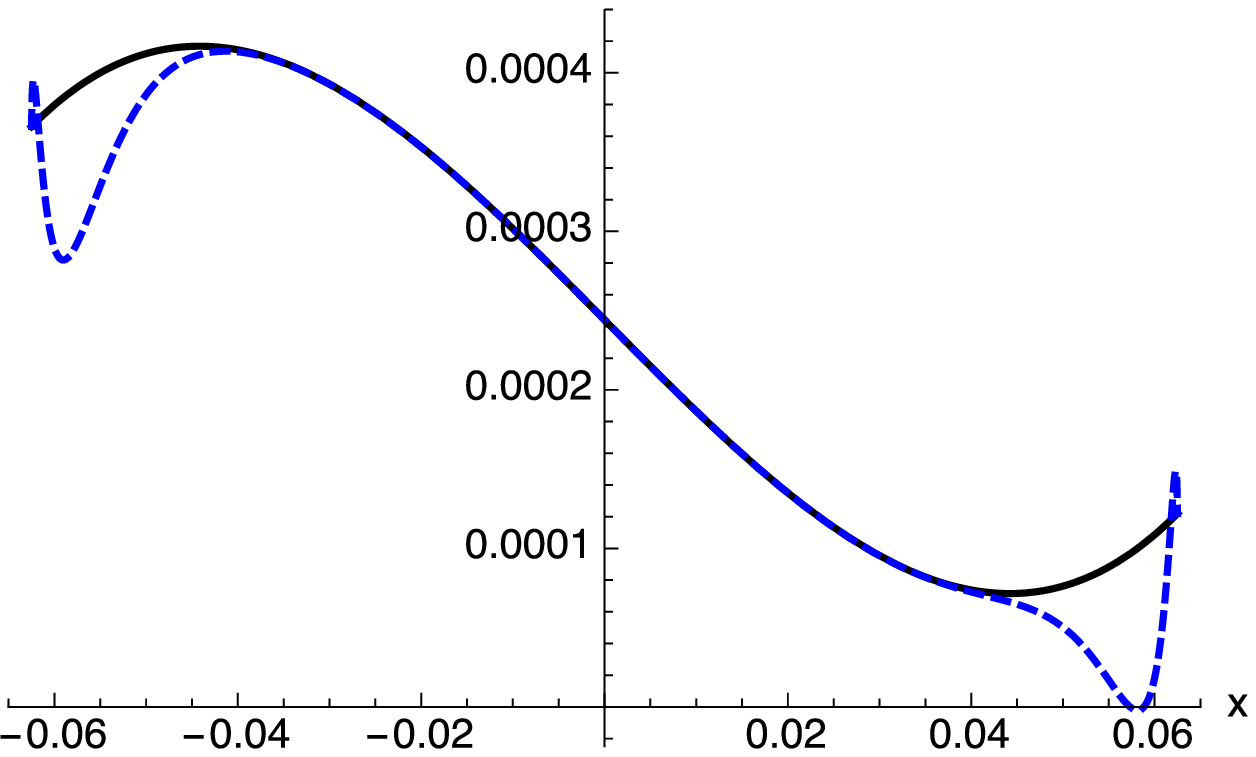}
\hfill
\includegraphics[width=0.45\textwidth]{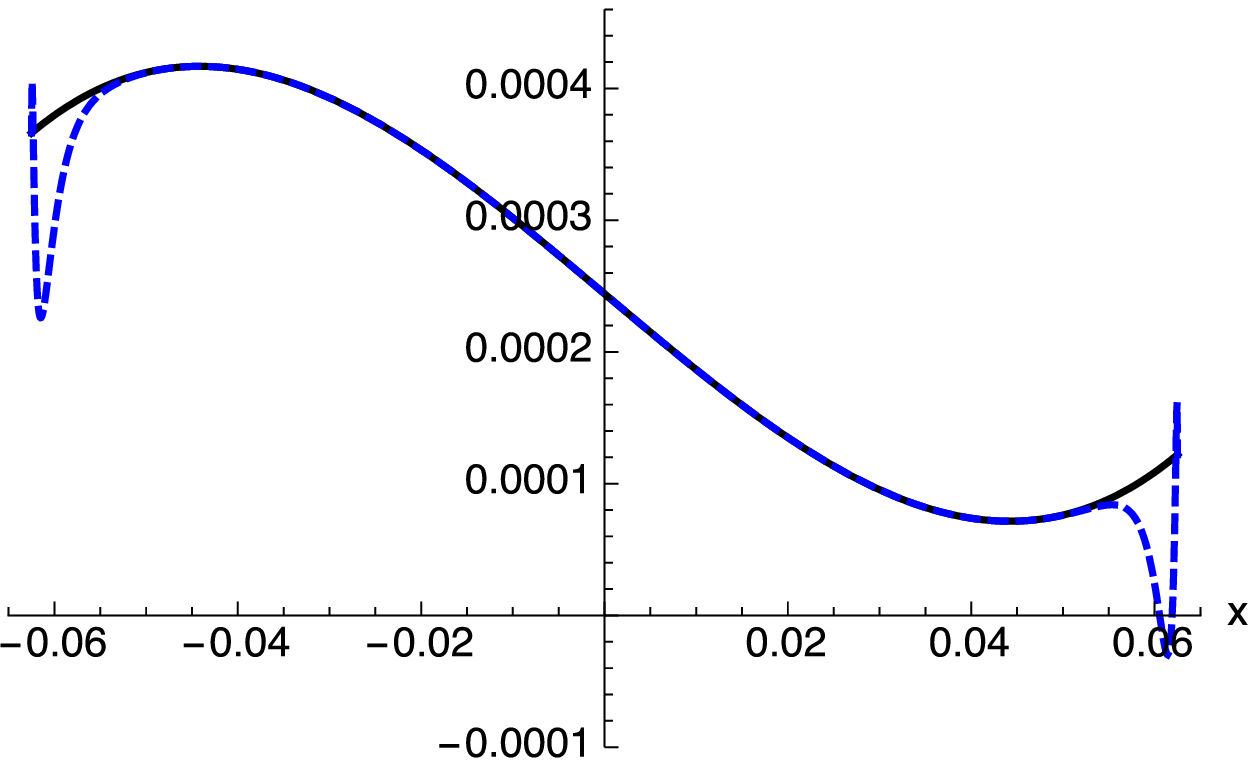}
\caption{In the commutative case (solid line) there are no critical points for $b < b_0 = 1/\sqrt{8} Q$ such that $|x| < b/4Q$; noncommutativity  modifies this behavior and critical points arise as a consequence ($\theta = 0.1$ (left figure) and $\theta = 0.05$ (right figure), dashed line).}  
\label{fig2}
\end{figure}

Fig.~\ref{fig2} shows the graphic solution to Eq.~(\ref{critpmodela}) when $Q=1,8\pi M = 1, b = 1/4$ for $\theta = 0.1$ and $\theta = 0.05$ respectively. In the commutative case, for the value of $b$ given above, the cubic polynomial on the left hand side of this equation has no real roots in the interval $|x| \leq b/4Q = 0.0625$; no critical points exist in this regime. The role of noncommutativity is to change this behavior, leading to critical points at $x = 0.057328$ and $x = 0.0588325$ for $\theta = 0.1$, where the dashed line crosses the $x$-axis (left figure), and to two critical points located at $x = 0.605505$ and $x = 0.0614104$ for $\theta = 0.05$ (right figure).  

\subsubsection{Gibbs energy}

\begin{widetext}

\begin{figure}[htbp]
\centering
\subfigure[] 
{
    \label{orbits:a}
    \includegraphics[width=0.3\textwidth]{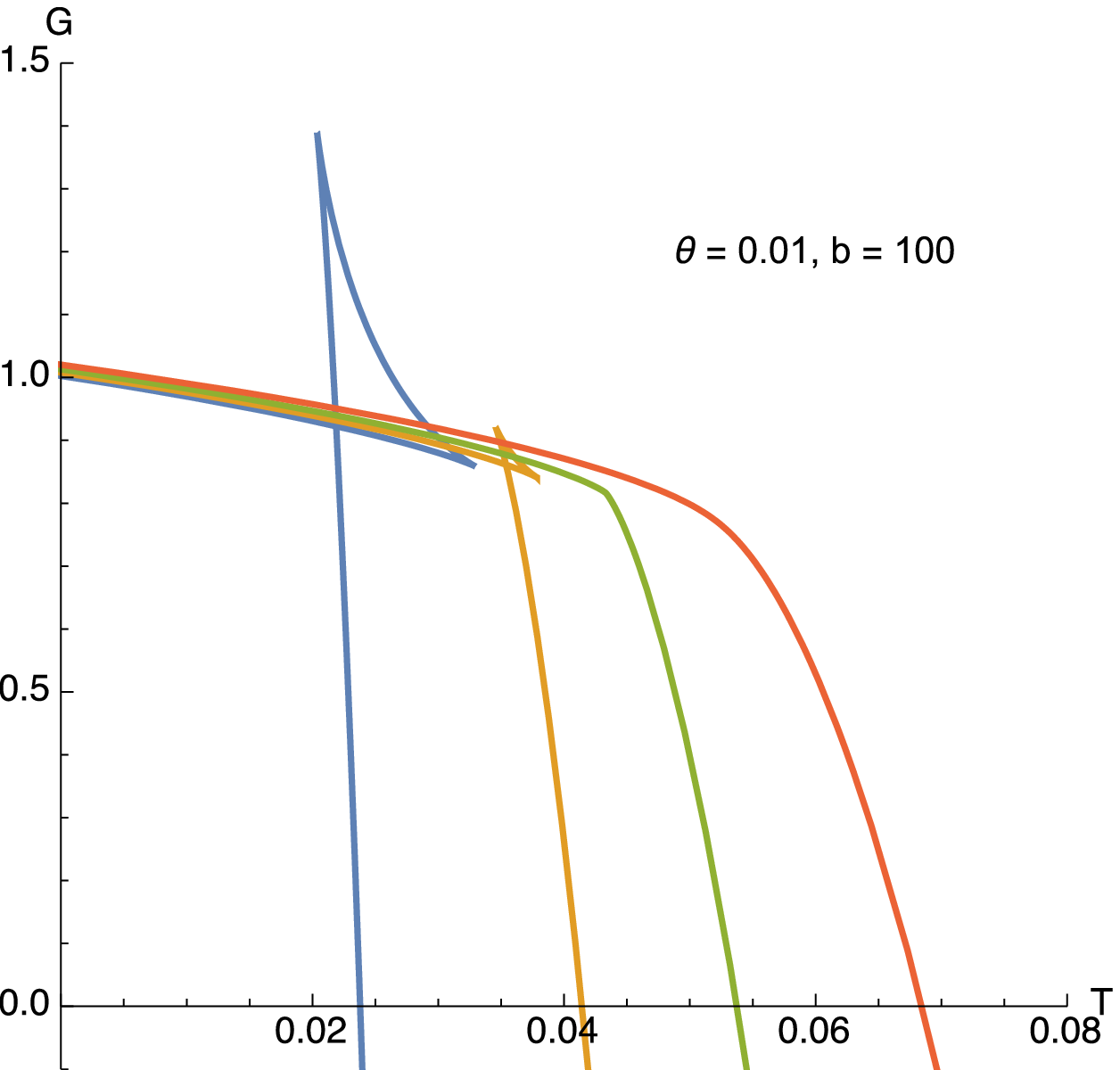}
}
\subfigure[] 
{
    \label{orbits:b}
    \includegraphics[width=0.3\textwidth]{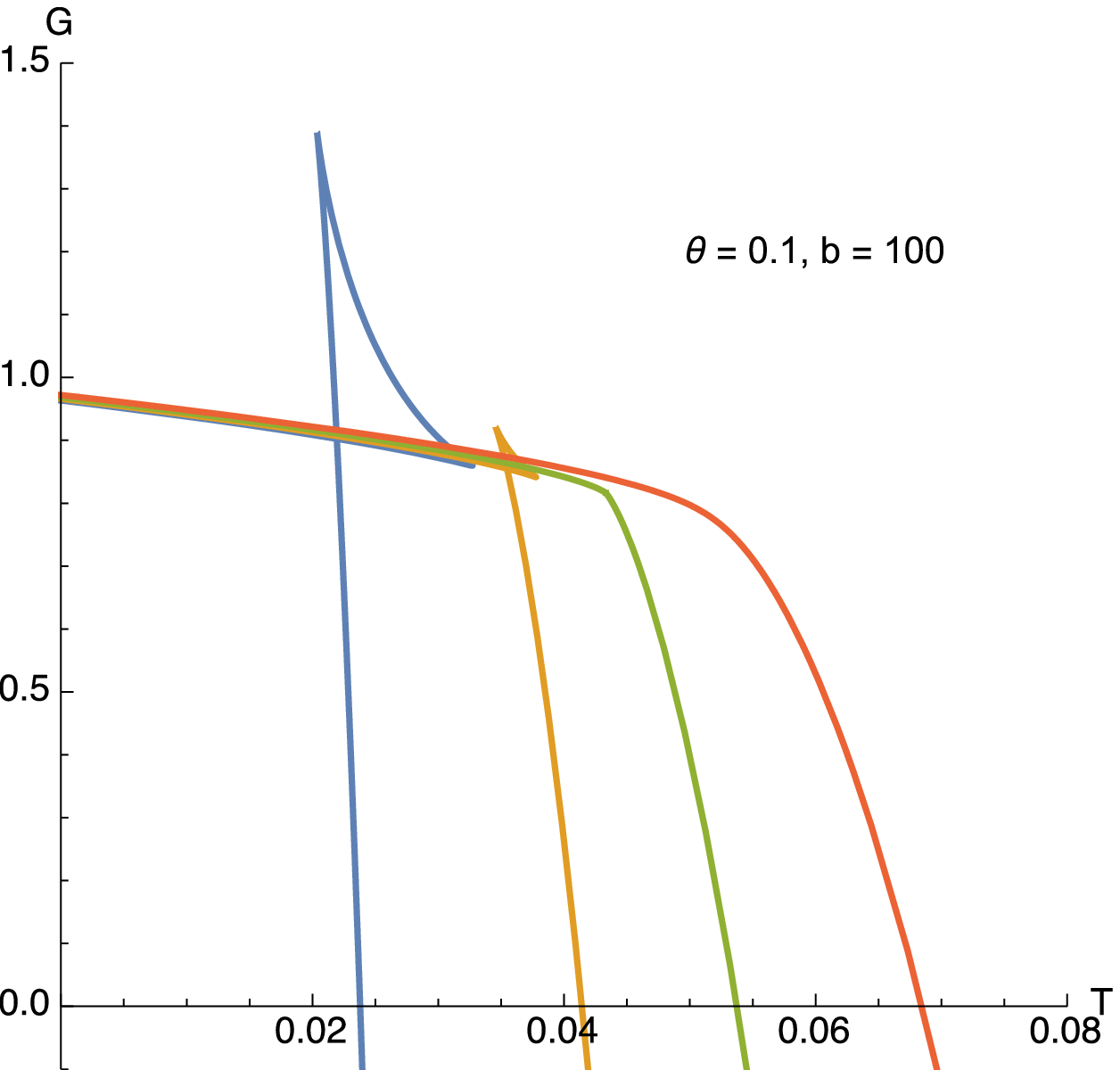}
}
\subfigure[] 
{
    \label{orbits:c}
    \includegraphics[width=0.3\textwidth]{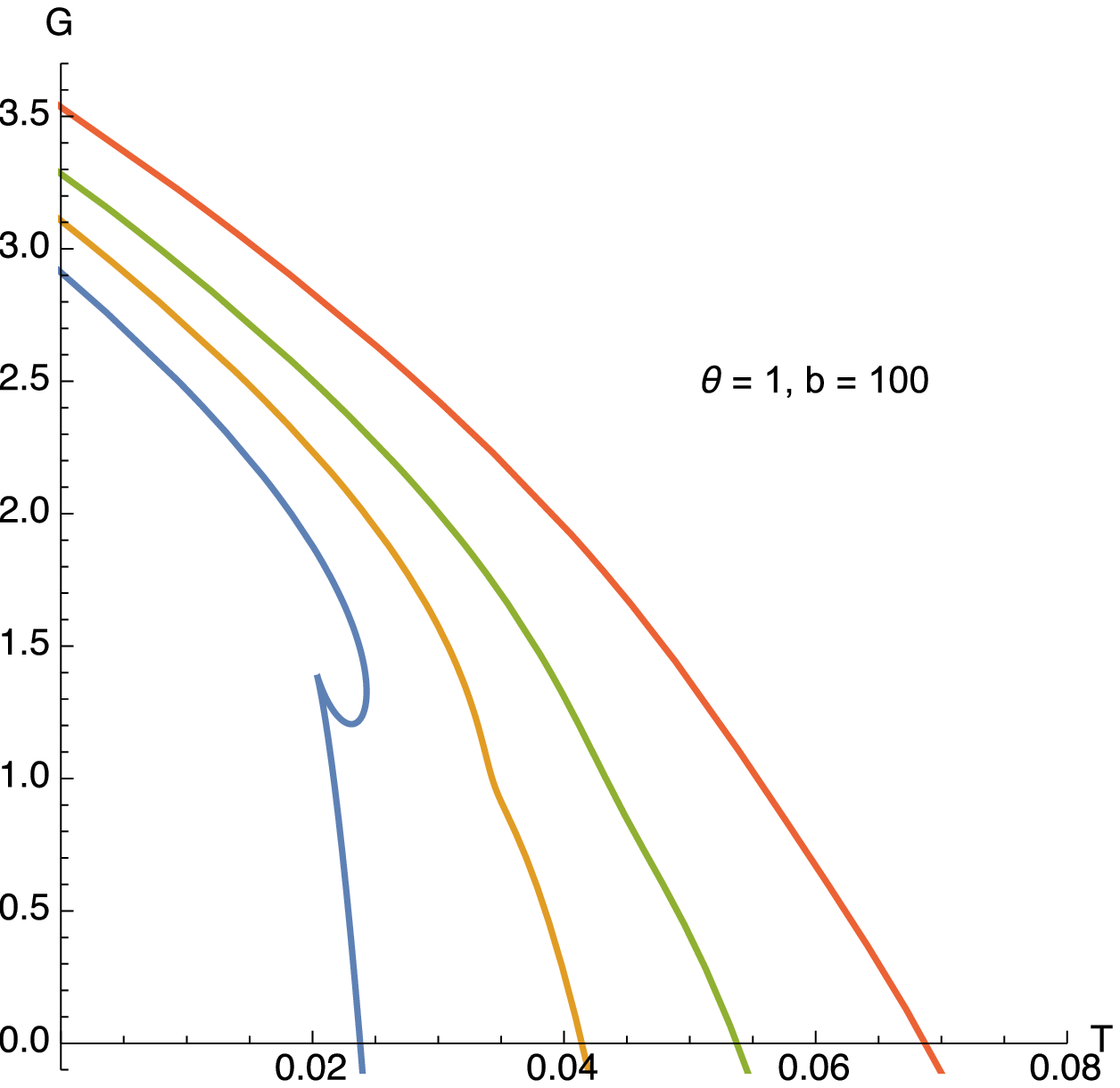}
}
\subfigure[] 
{
    \label{orbits:d}
    \includegraphics[width=0.3\textwidth]{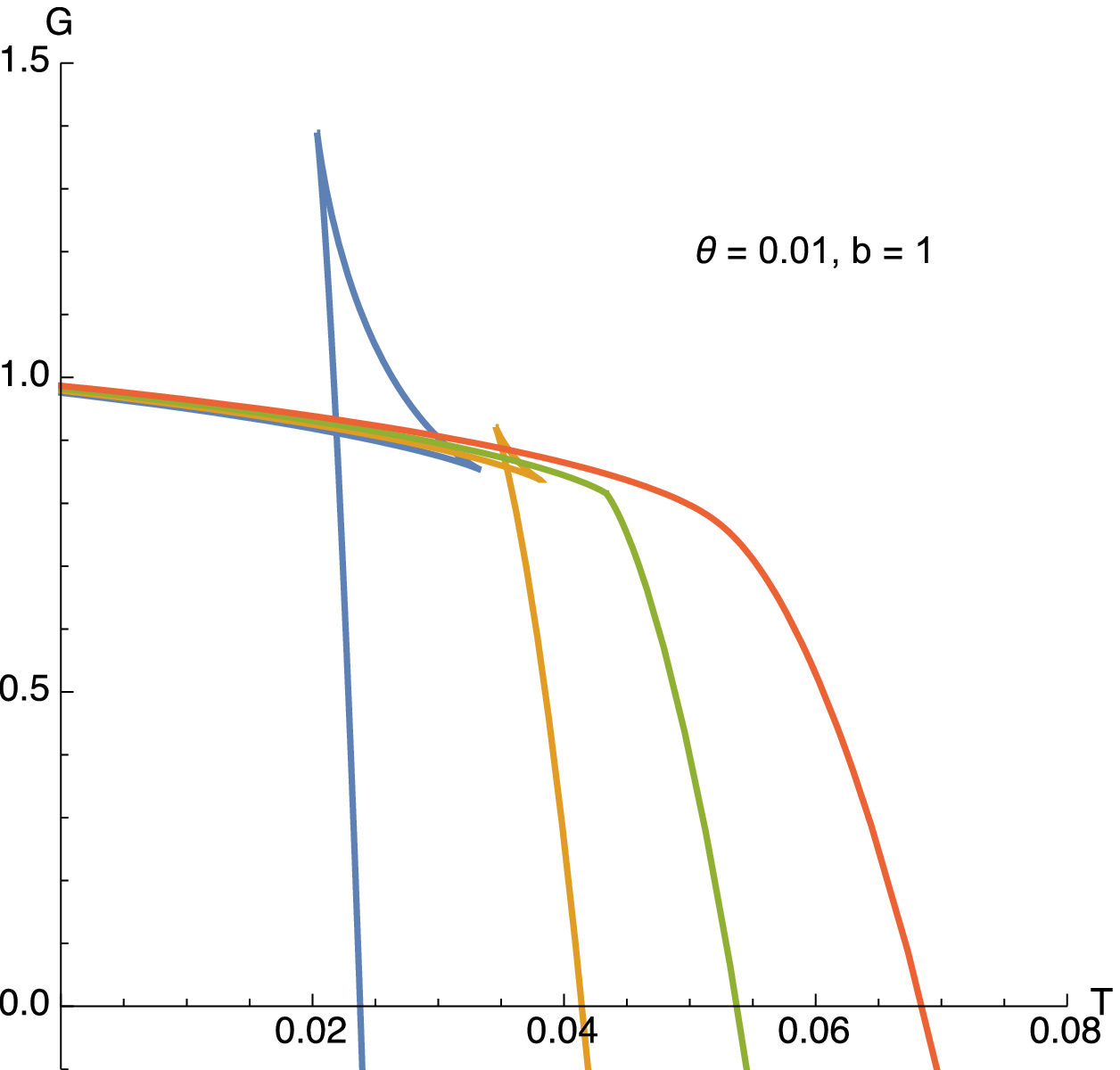}
}
\subfigure[] 
{
    \label{orbits:e}
    \includegraphics[width=0.3\textwidth]{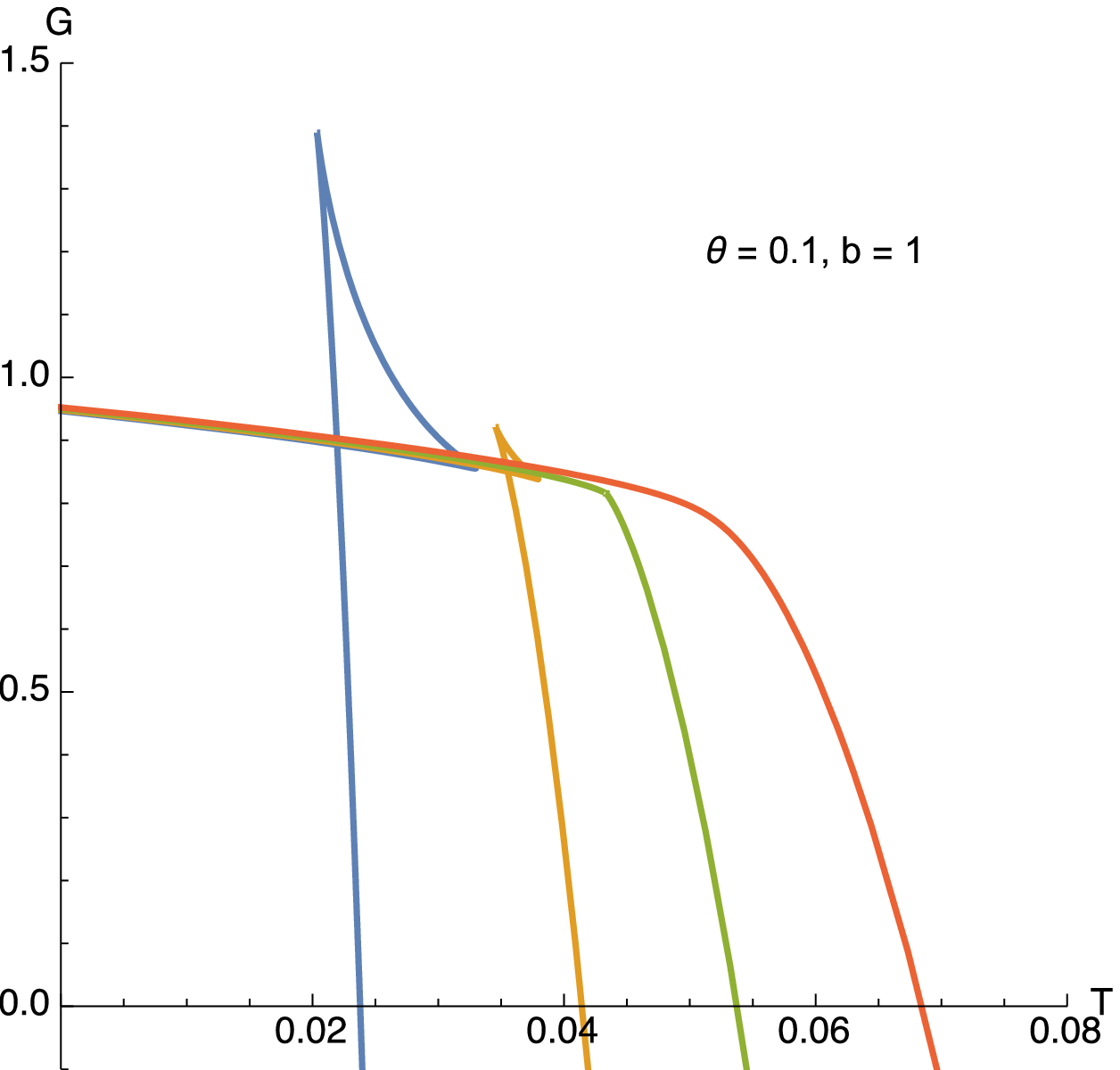}
}
\subfigure[] 
{
    \label{orbits:f}
    \includegraphics[width=0.3\textwidth]{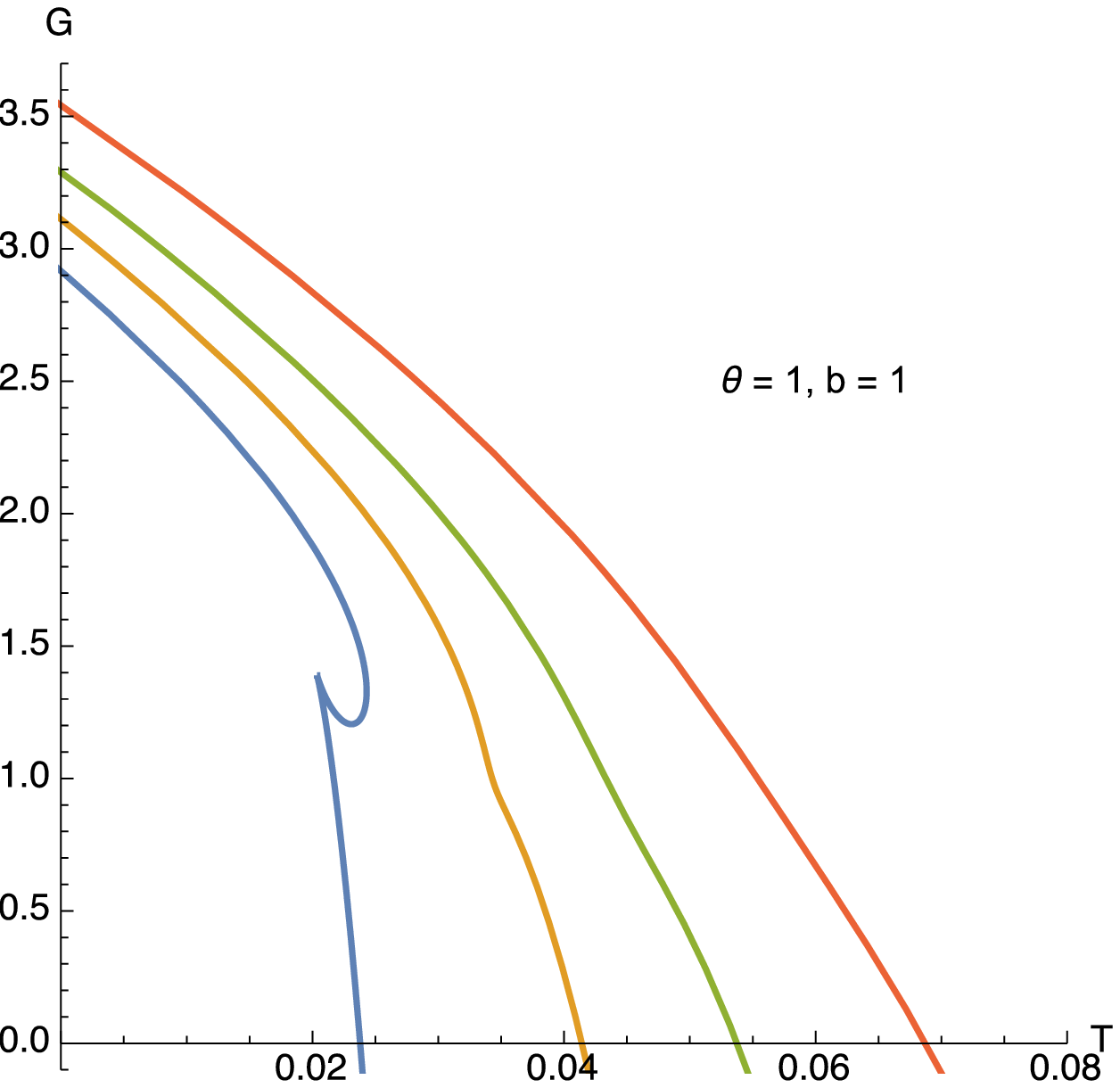}
}
\subfigure[] 
{
    \label{orbits:g}
    \includegraphics[width=0.3\textwidth]{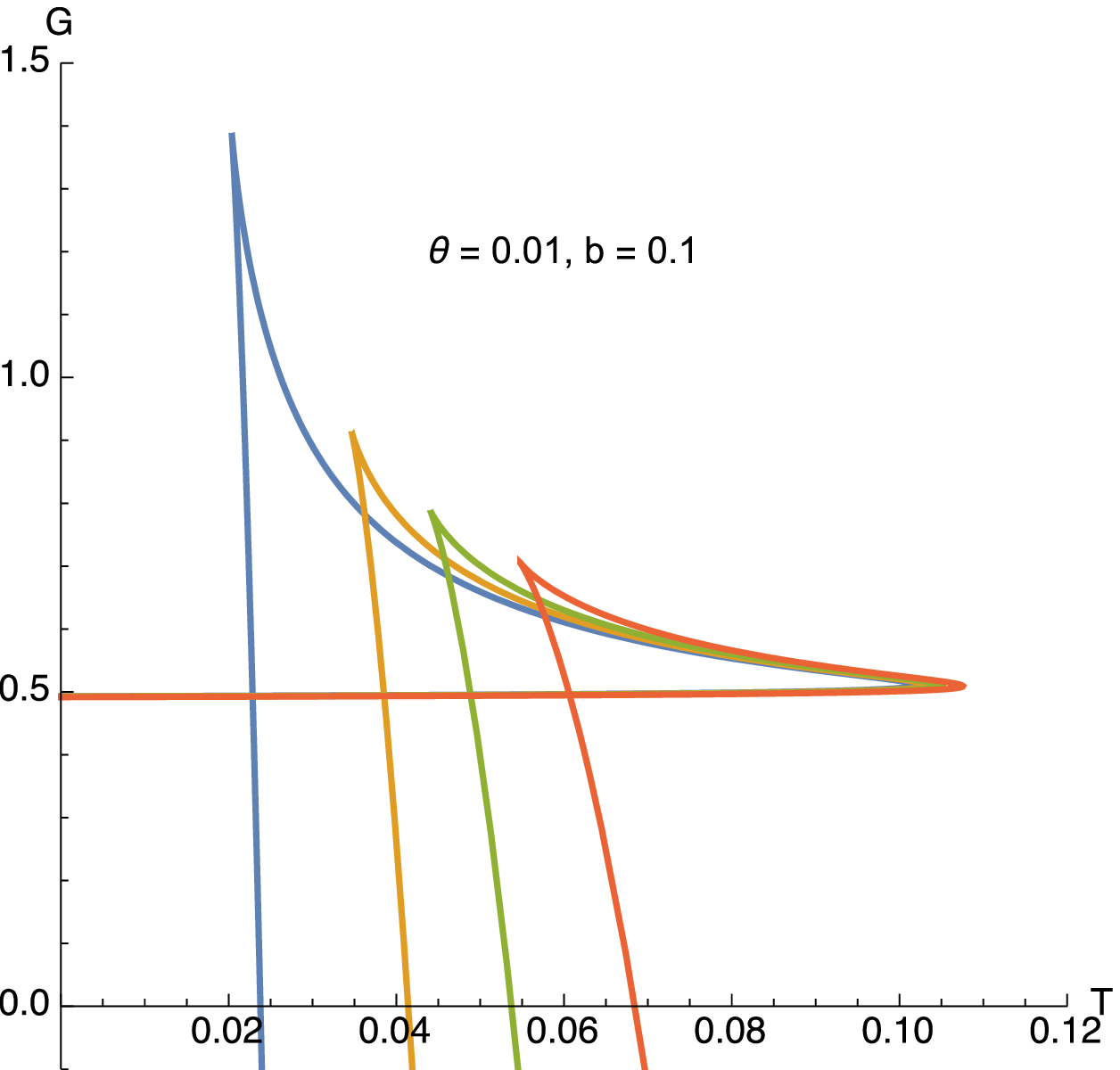}
}
\subfigure[] 
{
    \label{orbits:h}
    \includegraphics[width=0.3\textwidth]{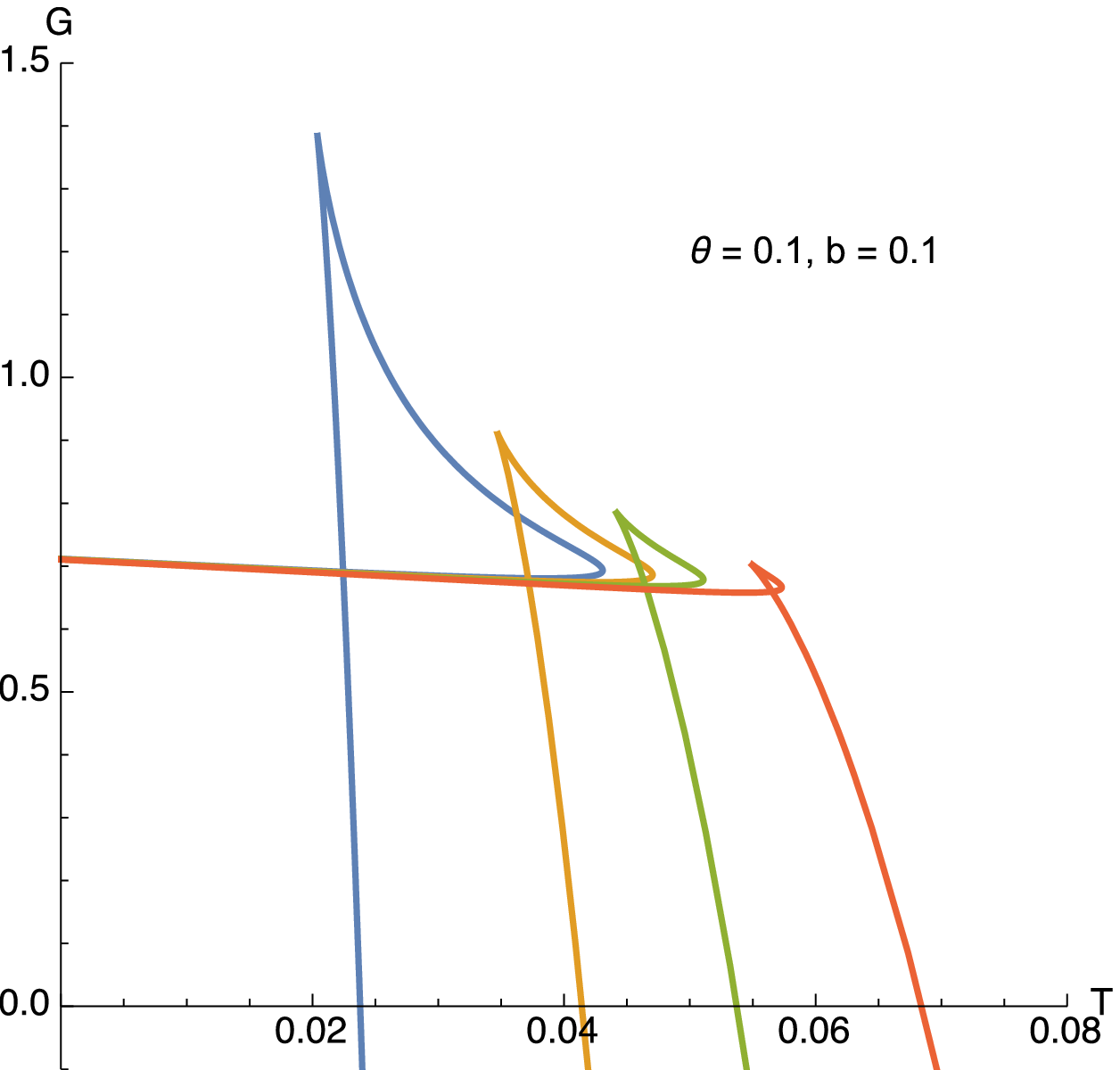}
}
\subfigure[] 
{
    \label{orbits:i}
    \includegraphics[width=0.3\textwidth]{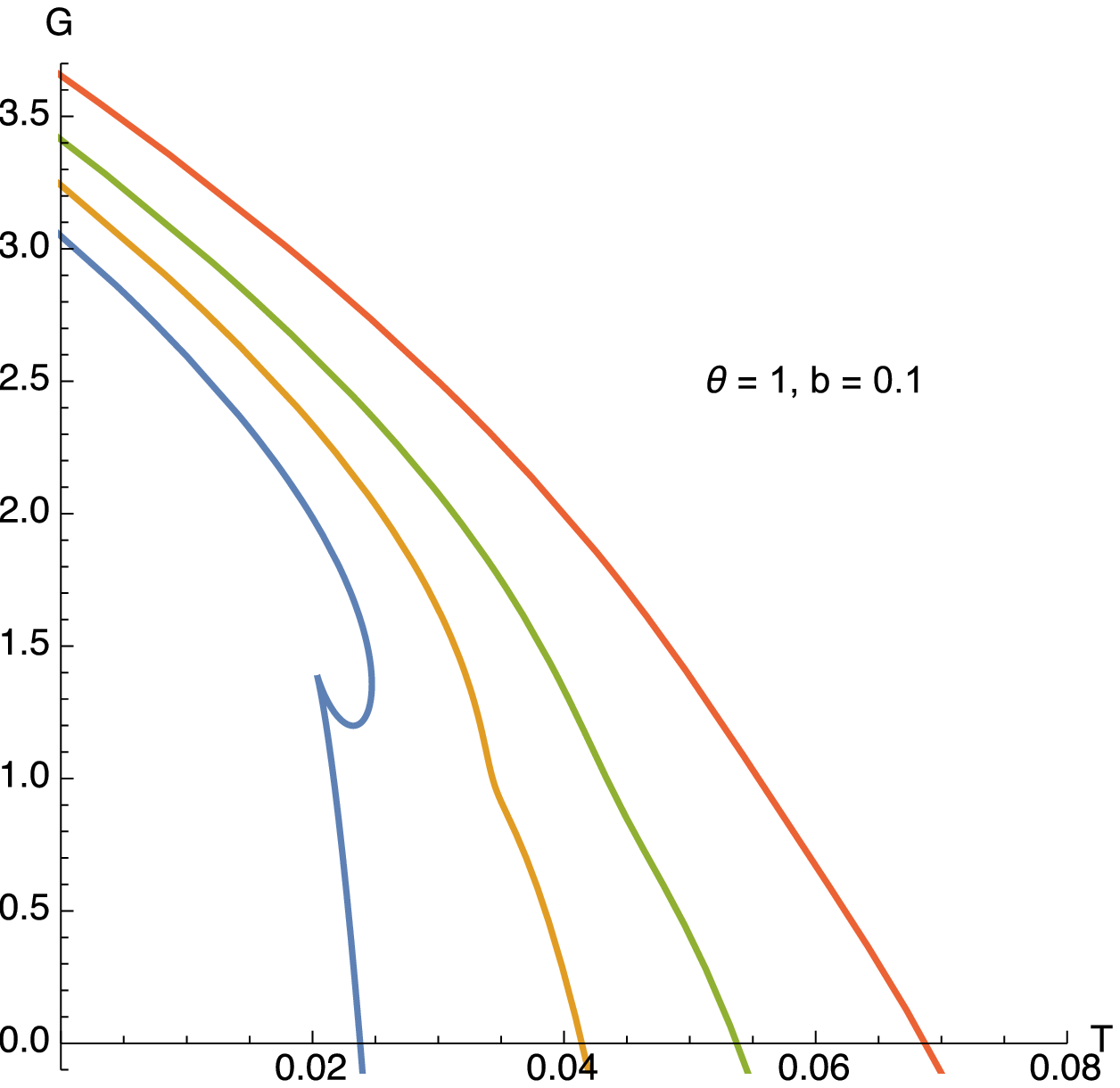}
}
\caption{The plots show the Gibbs function for model A for different values of the BI parameter $b$ and the noncommutative parameter $\theta$, as a function of temperature for fixed pressure and $Q = 1$. In Fig.~\ref{orbits:a} the known behavior for the Reissner-Nördstrom case is seen meanwhile Fig.~\ref{orbits:g} shows the typical behavior for the Schwarzschild solution.}
\label{fig3} 
\end{figure}
\end{widetext}

The Gibbs energy can be readily be found using the relationship
\be
G = \beta (M_{ADM} - T S) = \beta (M - T S),
\ee
where $M$ is the mass of the source written in terms of the horizon radius $r_H$ (see Eq.~(\ref{massatrh})), $T$ the temperature of the BI black hole at $r_H$, $S$ its entropy and $\beta$ a constant. Knowledge of the Gibbs function $G$ will allow us to infer the presence of phase transitions. Explicitly we have in (3+1)-dimensions

\begin{eqnarray}
G &=& \beta \left( M \left[ 1 - \frac {r_H}{4(4\pi\theta)^{3/2}} e^{-r_H^2/4\theta} \right] - \frac 14 \left[ r_H \right. \right.
\nonumber \\[4pt]
&&\left. \left. - \Lambda r_H^3 + 2b^2 r_H^3 \left( 1 - \sqrt{1+ \frac {Q^2}{b^2 r_H ^4}} \right) \right] \right),
\label{gibbsmodela}
\end{eqnarray}
where $M$, as a function of $r_H$, is given in Eq.~(\ref{massatrh}). In Fig.~\ref{fig3} we show the Gibbs energy for different values of the parameters $b$ and $\theta$. It can be noticed that noncommutativity changes the location of the points where first order phase transitions happen.

\subsection{Model B}
\label{subsec4}

As in the previous section, we have the expression for the thermodynamic internal energy
\be
E(R) = - R N(R) - 4\pi R^2\epsilon_0(R),
\ee
together with the pressure
\be
{\cal P} = \frac {\p}{\p (4\pi R^2)} (R N(R)) + \frac {\p (R^2 \epsilon_0 (R))}{\p R^2},
\ee
where $\epsilon_0(R)$ is given by Eq.~(\ref{epsilon0}). We now follow a similar procedure as before: first, the lapse function $N^2(r)$ is given by the right hand side of Eq.~(\ref{lapseb}) and therefore, the mass $M$ is written as
\begin{widetext}
\be
M = \frac {r_H^{d-2}}2\left[ 1 - \frac 1d \Lambda r_H^2 + \frac {2b^2 r_H^2}d \left( 1- \sqrt{1 + \frac {H^2(r_H)}{b^2 r_H^{2(d-1)}}}\right) + \frac 2d \frac 1{r_H^{d-2}} \int_{r_H}^\infty ds \frac { - H(s) s^d \rho_Q(s) + (d -1) H^2(s)}{\sqrt{s^{2(d-1)} + H^2(s)/b^2}} \right],
\ee
using as before the condition $N(r_H) = 0$. The surface gravity, as deduced from Eq.~(\ref{diffequmodelb}), is given by
\be
\kappa_H = \frac 1{2r_H^{d-2}} \left[ (d - 2) r_H^{d-3} - \Lambda r_H^{d-1} + 2b^2 r_H^{d-1} \left( 1 - \sqrt{1 + \frac {H^2(r_H)}{b^2 r_H^{2(d-1)}}} \right) \right].
\ee
For the (3+1)-dimensional case, we have that the temperature at the hypersurface $r = R = const.$ is
\begin{eqnarray}
T &=& \frac 1{N(R)} \frac 1{4\pi r_H} \left[ 1 - \Lambda r_H^2 + 2b^2 r_H^2 \left( 1 - \sqrt{1 + \frac {H^2(r_H)}{b^2 r_H^4}} \right) \right],
\label{temp1modelb}
\end{eqnarray} 
where
\be
N^2(R) =  1 - \frac {2M}R - \frac 13 \Lambda R^2 + \frac 23 b^2 R^2 \left( 1- \sqrt{1 + \frac {H^2(R)}{b^2 R^4}} \right) + \frac 23 \frac 1R \int_R^\infty ds \frac { - H(s) s^3 \rho_Q(s) + 2 H^2(s)}{\sqrt{s^4 + H^2(s)/b^2}},
\ee
and
\be
M = \frac {r_H}2\left[ 1 - \frac 13 \Lambda r_H^2 + \frac 23 b^2 r_H^2 \left( 1- \sqrt{1 + \frac {H^2(r_H)}{b^2 r_H^4}} \right) + \frac 23 \frac 1{r_H} \int_{r_H}^\infty ds \frac { - H(s) s^3 \rho_Q(s) + 2 H^2(s)}{\sqrt{s^4 + H^2(s)/b^2}} \right].
\ee
\end{widetext}

\subsubsection{Equation of state}

\begin{figure}[hbtp]
\centering
\includegraphics[width=0.45\textwidth]{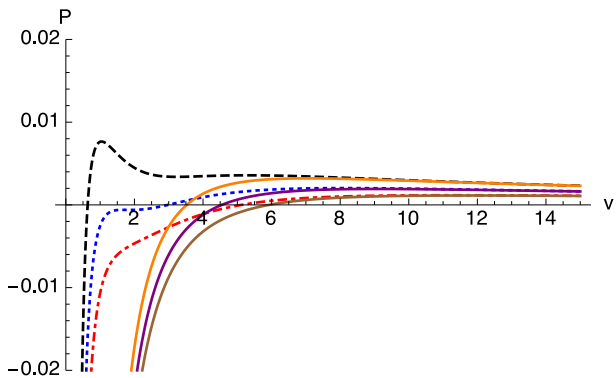}
\caption{$P-v$ diagram of noncommutative BI black hole (charge deformation). The thick lines correspond to the noncommutative equation of state with $\theta = 0.1$ for $T = 0.04517, 0.035$ and $0.026885$ (top to bottom); the dashed lines are the corresponding commutative cases. We have set $Q=10$ and $b = 0.45$.}  
\label{fig4}
\end{figure}

We now obtain the equation of state for this solution in the (3+1)-dimensional case; we again follow~\cite{Gunasekaran:2012dq}. As the temperature we take again $T = \kappa_H/2\pi$; with the same definitions of pressure $P = -\Lambda/8\pi$ and specific volume $v = 2 r_H$ as before, we have
\be
P = \frac Tv - \frac 1{2\pi v^2} - \frac {b^2}{4\pi} \left( 1 - \sqrt{1 + \frac {16 H^2(v)}{b^2 v^4}} \right).
\ee
This expression is similar to the standard result but for the function $H(v)$ that plays the role of the charge $Q$. As seen in Fig.~\ref{fig4}, in this model noncommutativity has a more visible effect on the classical $P-v$ diagram.

Critical points are defined again by the conditions
\be
\left( \frac {\p P}{\p v} \right)_{T_c} = 0, \qquad \left( \frac {\p^2 P}{\p v^2} \right)_{T_c} = 0.
\ee
We define a new variable
\begin{equation}
x^{-2} := v^4 + \frac {16 H^2(v)}{b^2},
\label{xvmodelb}
\end{equation}
and after a long but straightforward calculation we obtain
\be
x^3 + p(v) x + q(v) = 0,
\label{cubicmodelb}
\ee
as the equation determining the critical points of the model. We have again a generalization of the cubic algebraic relation for the variable $x$. Here the functions $p(v)$ and $q(v)$ are given by
\begin{eqnarray}
s(v)p(v) &:=& -{b^2\over 4}[v^2{H^\prime}^2(v/2)+v^2H(v/2)H^{\prime\prime}(v/2)
\nonumber \\[4pt]
&&-6vH(v/2)H^\prime(v/2)+6H^2(v/2)],
\nonumber \\[4pt]
s(v)q(v) &:=& \frac {b^2}{16},
\nonumber \\[4pt]
s(v) &:=& [2vH(v/2)H^\prime(v/2)-4H^{2}(v/2)]^2,
\end{eqnarray}
and it is understood that the variable $v$ in these expressions is defined implicitly in terms of $x$ through the relation Eq.~(\ref{xvmodelb}). Here $H^\prime$ denotes derivative of $H$ with respect to its argument.

\begin{figure}[h]
\centering
\includegraphics[width=0.45\textwidth]{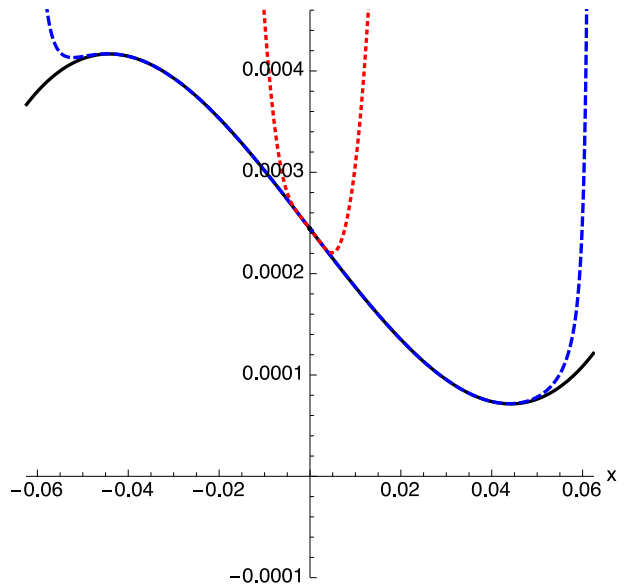}
\caption{In the commutative case (solid line) there are no critical points for $b < b_0 = 1/\sqrt{8} Q$ such that $|x| < b/4Q$; in the noncommutative model B this feature is still present: the plot shows the behavior for $\theta = 0.1$ (dashed line) and $\theta = 2$ (dotted line).}  
\label{fig5}
\end{figure}

It can be seen that in the limit $\theta \to 0$ the functions $p(v)$ and $q(v)$ reduce to the standard expressions $-3b^2/32Q^2$ and $3b^2/256Q^2$ respectively. 
In Fig.~\ref{fig5} we have plotted the curve given by Eq.~(\ref{cubicmodelb}) for $\theta = 0.1$ and $\theta = 0.2$. We see that in this case noncommutativity does not lead to the appearance of critical points in the region $|x| < b/4Q$.

\subsubsection{Gibbs energy}

We use here again the expression
\be
G = \beta (M_{ADM} - T S) = \beta (M - TS),
\ee
to find the Gibbs energy. We have
\begin{widetext}
\be
G = \frac \beta4\left[ r_H + \frac 13 \Lambda r_H^3 - \frac 23 b^2 r_H^3 \left( 1- \sqrt{1 + \frac {H^2(r_H)}{b^2 r_H^4}} \right) + \frac 43 \int_{r_H}^\infty ds \frac { - H(s) s^3 \rho_Q(s) + 2 H^2(s)}{\sqrt{s^4 + H^2(s)/b^2}} \right].
\label{gibbsmodelb}
\ee

In Fig.~\ref{fig6} we show the typical behavior of this function for different values of the parameters; it is very similar to that of the Schwarzschild solution even when noncommutativity is present.

\begin{figure}[htbp]
\centering
\subfigure[] 
{
    \label{orbits:aa}
    \includegraphics[width=0.3\textwidth]{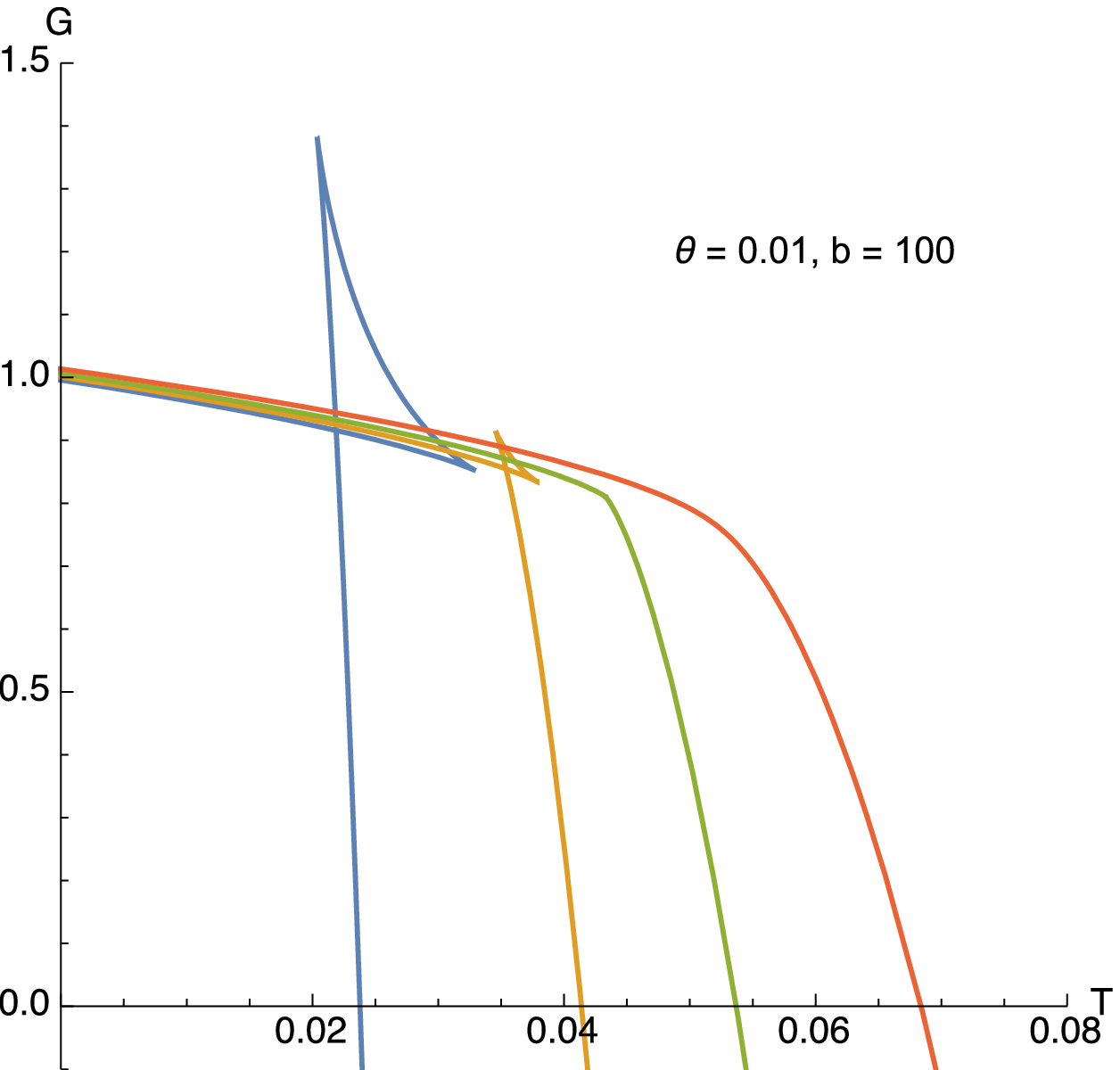}
}
\subfigure[] 
{
    \label{orbits:bb}
    \includegraphics[width=0.3\textwidth]{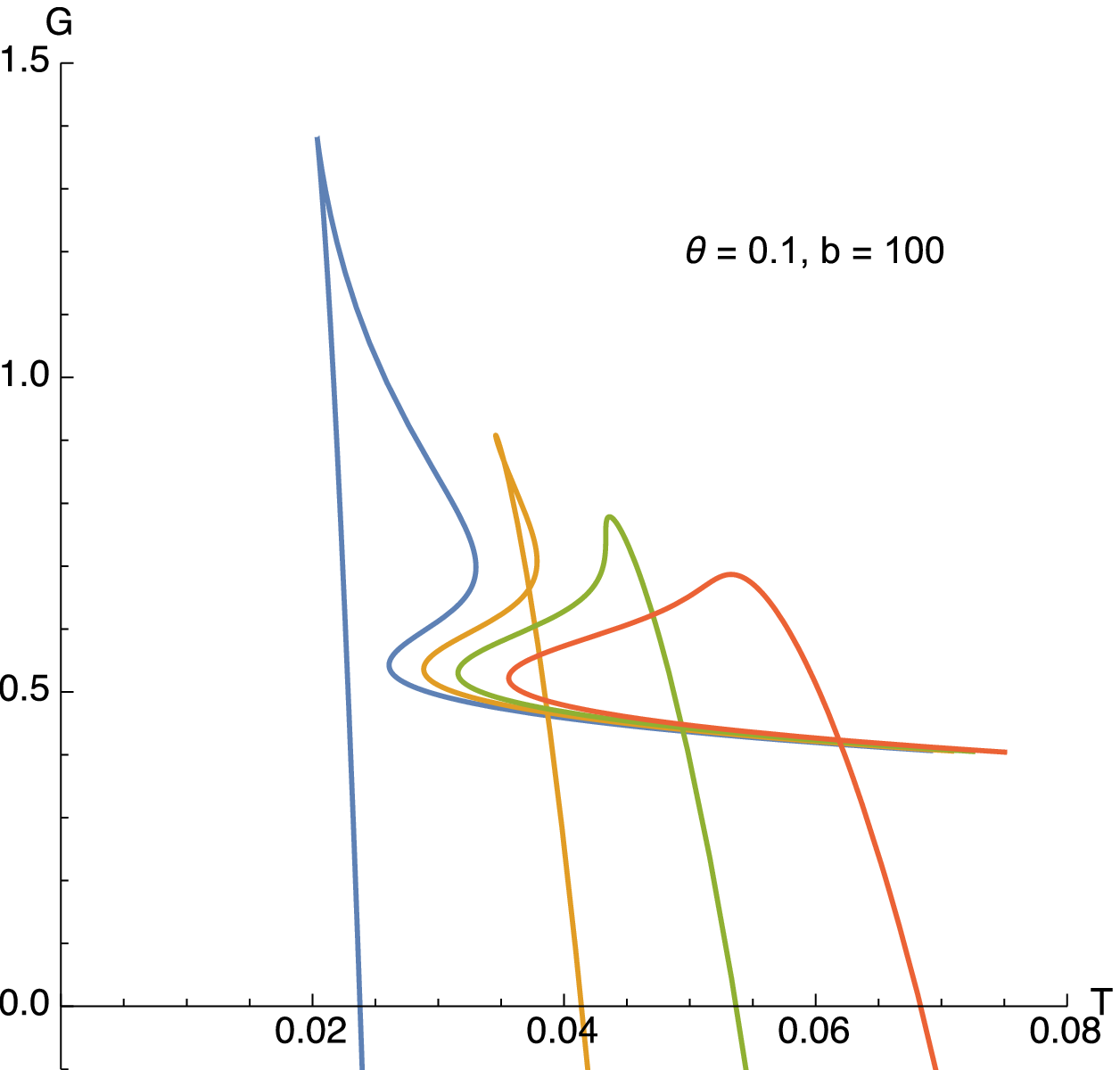}
}
\subfigure[] 
{
    \label{orbits:cc}
    \includegraphics[width=0.3\textwidth]{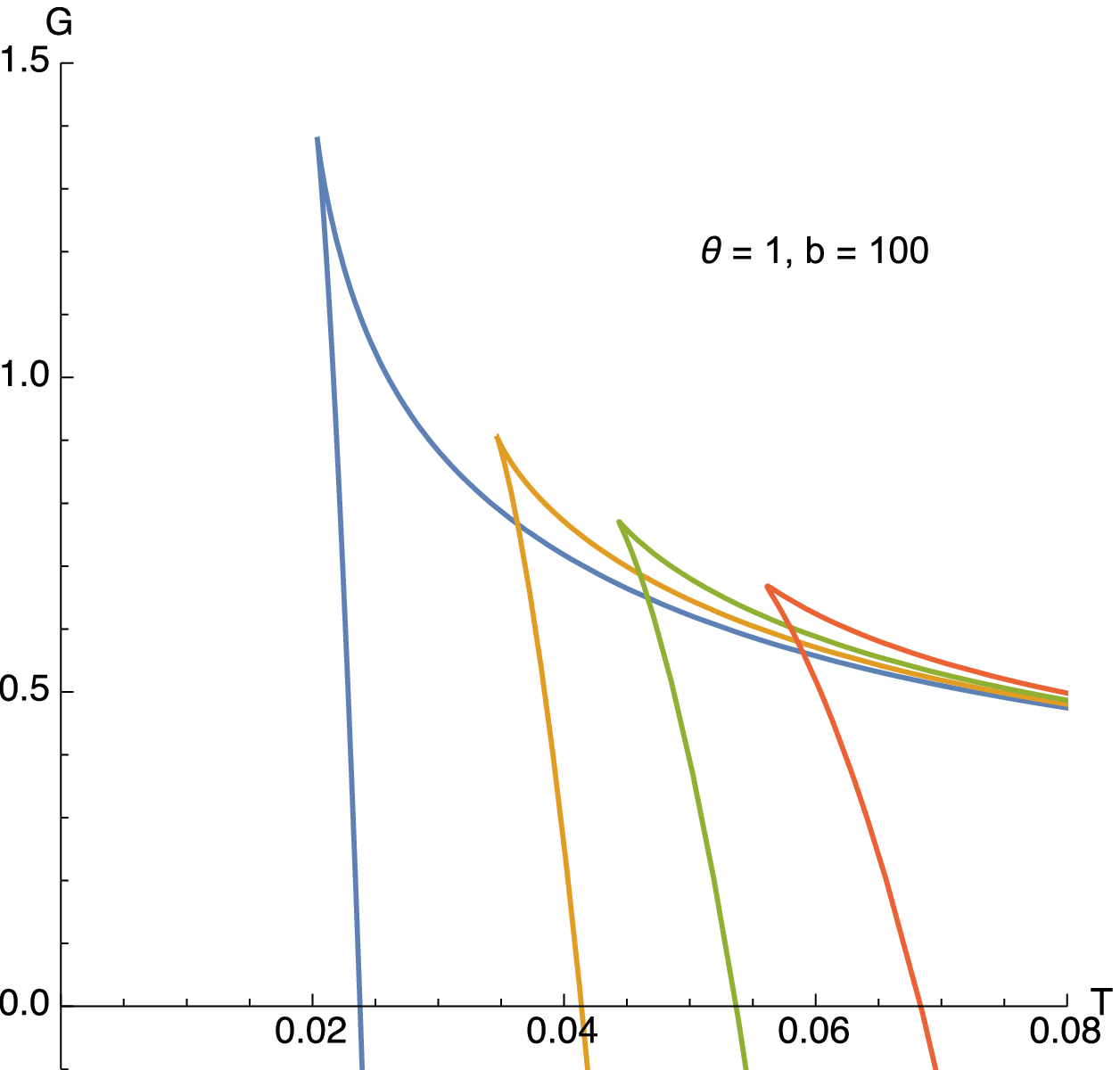}
}
\subfigure[] 
{
    \label{orbits:dd}
    \includegraphics[width=0.3\textwidth]{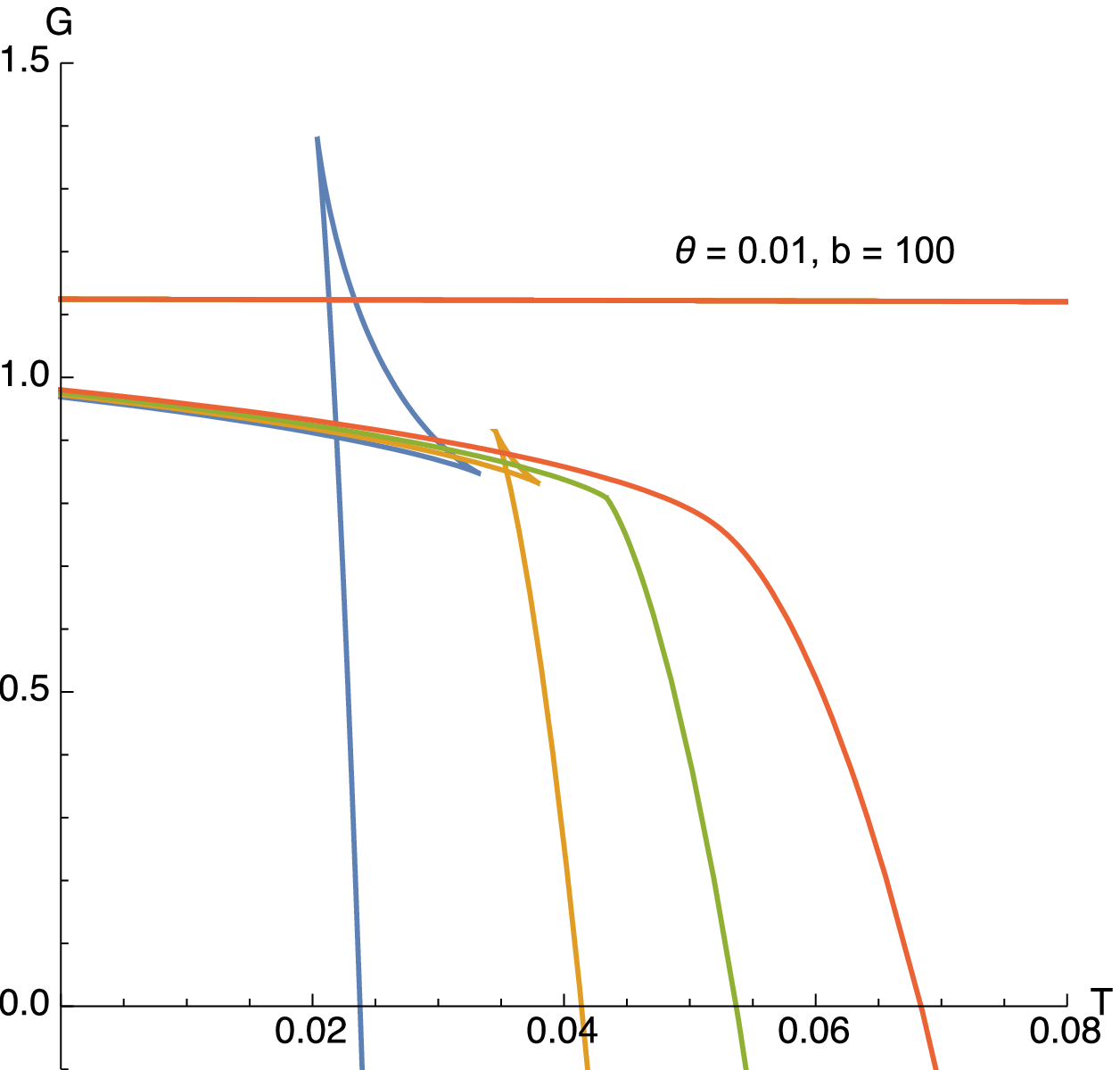}
}
\subfigure[] 
{
    \label{orbits:ee}
    \includegraphics[width=0.3\textwidth]{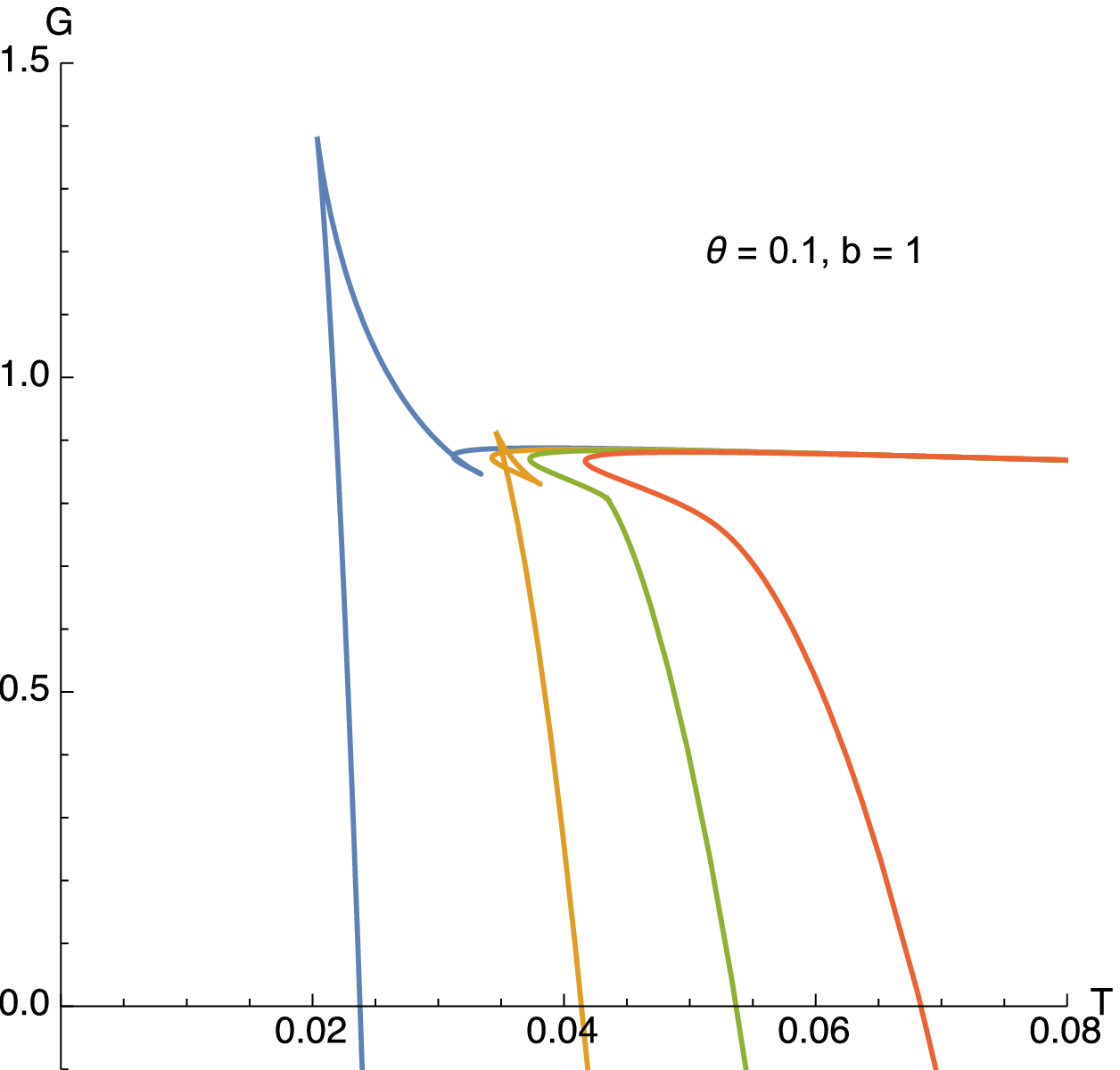}
}
\subfigure[] 
{
    \label{orbits:ff}
    \includegraphics[width=0.3\textwidth]{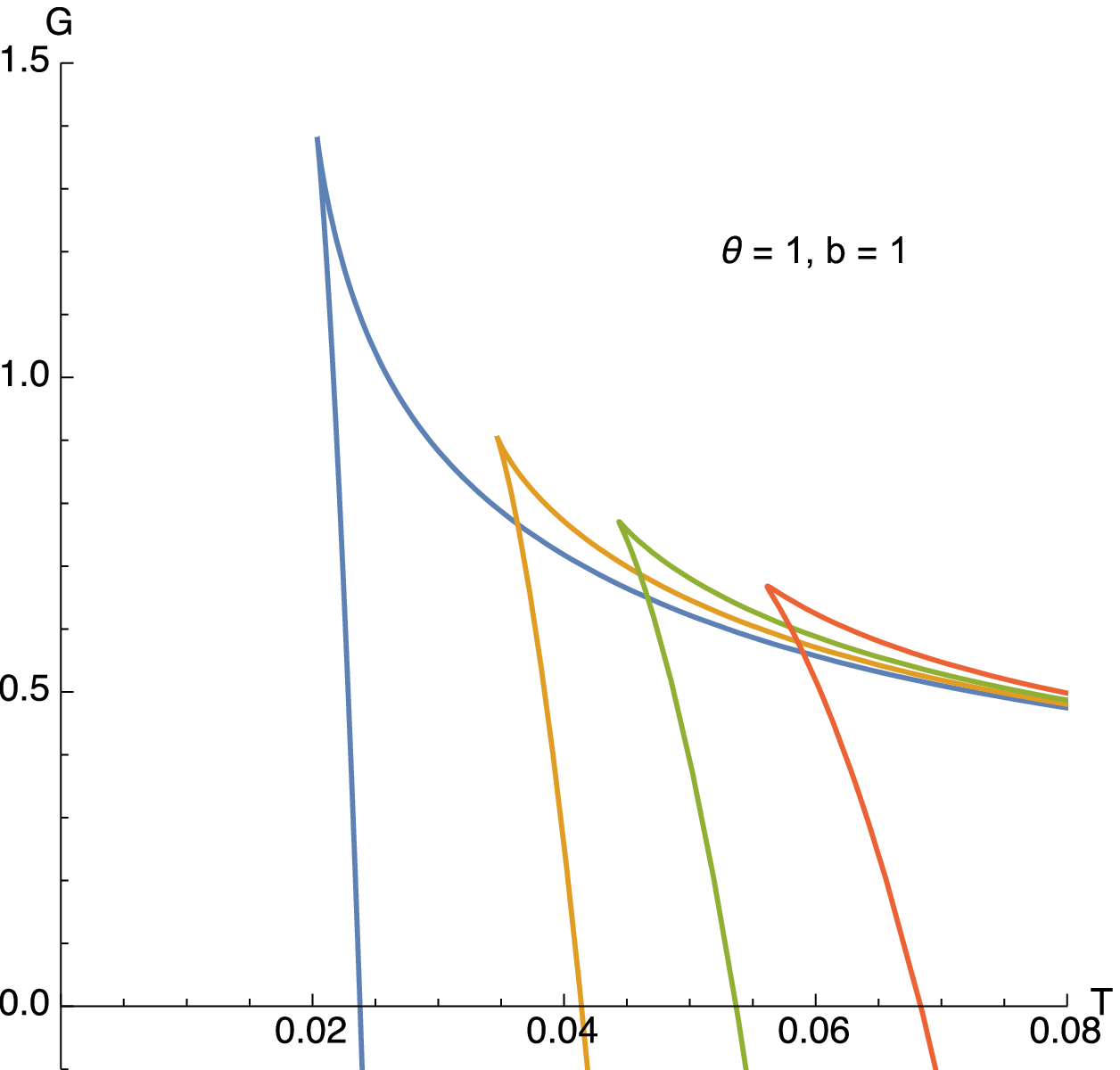}
}
\subfigure[] 
{
    \label{orbits:gg}
    \includegraphics[width=0.3\textwidth]{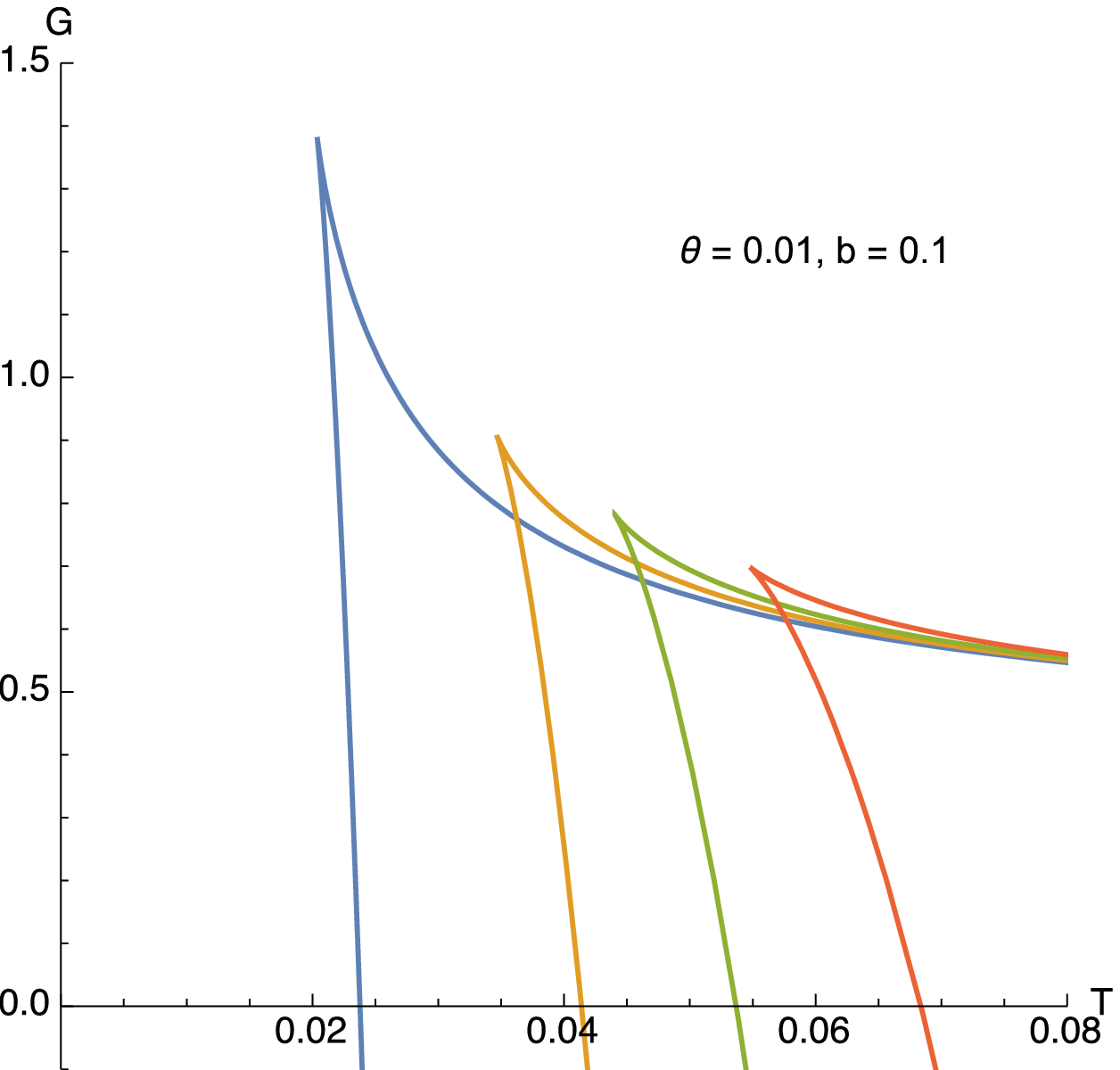}
}
\subfigure[] 
{
    \label{orbits:hh}
    \includegraphics[width=0.3\textwidth]{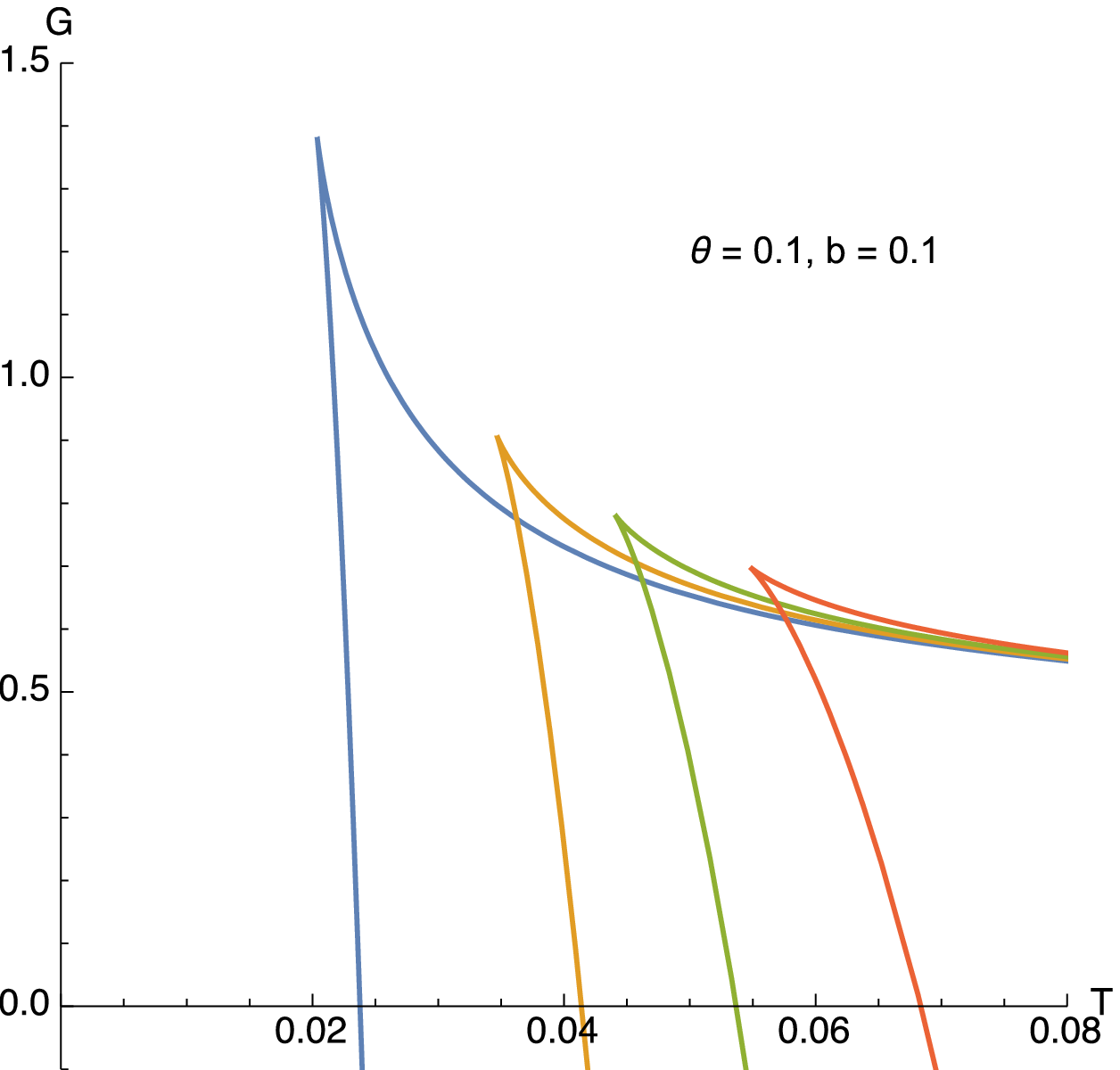}
}
\subfigure[] 
{
    \label{orbits:ii}
    \includegraphics[width=0.3\textwidth]{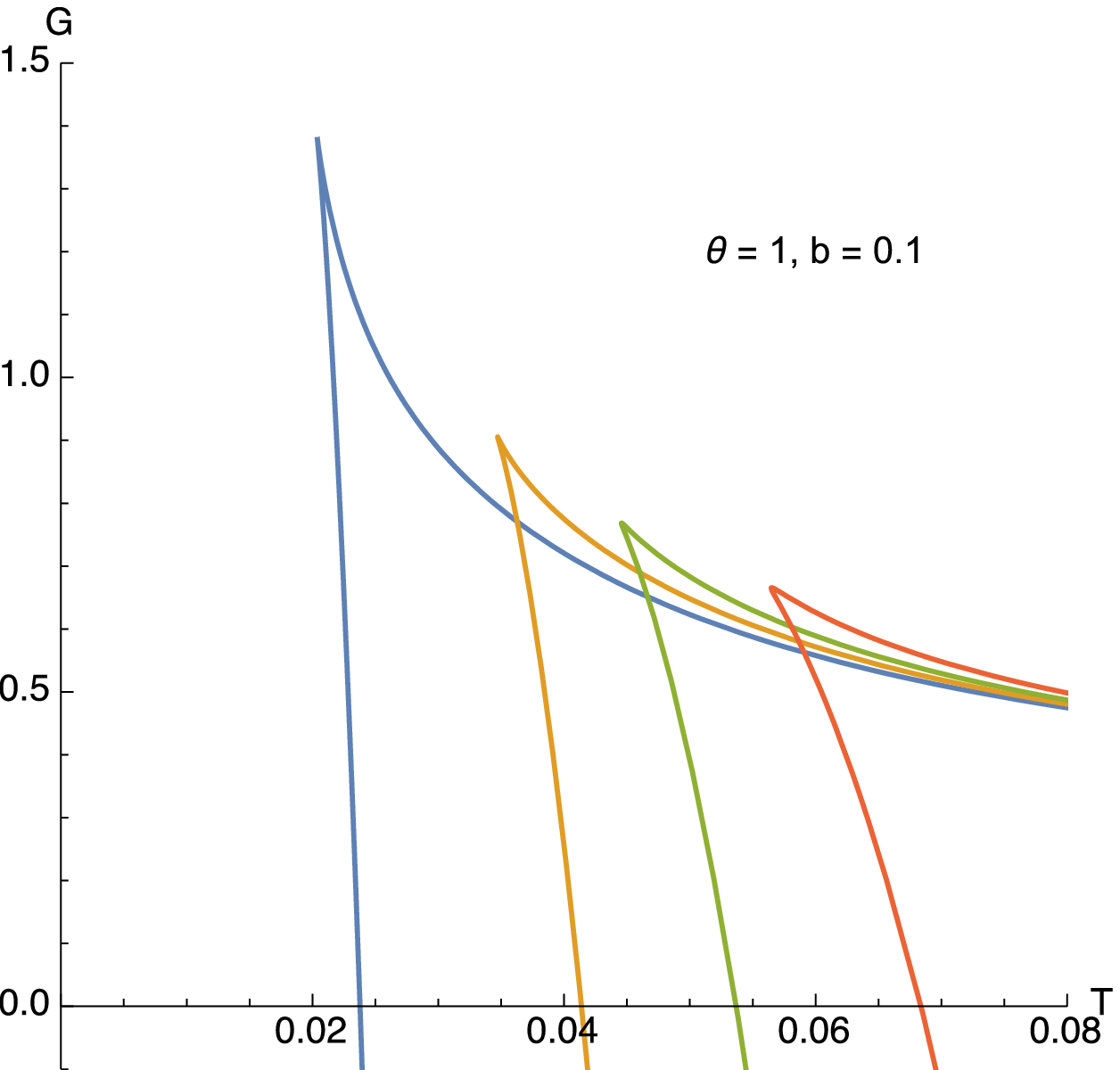}
}
\caption{The plots show the Gibbs fucntion for model B for different values of the BI parameter $b$ and the noncommutative parameter $\theta$, as a function of temperature for fixed pressure and $Q = 1$. Fig.~\ref{orbits:aa} reproduces the known behavior for the Reissner-Nördstrom case meanwhile in Fig.~\ref{orbits:gg} the typical behavior for the Schwarzschild solution is shown. In this case, for $b$ fixed, noncommutativity can eliminate some critical points.}
\label{fig6} 
\end{figure}
\end{widetext}

\section{Conclusions}

We have analyzed the influence of noncommutativity in the thermodynamic properties of the BI black hole. As a first step, we obtained the generic expressions for the mass and the temperature as a function of the horizon radius for arbitrary dimensions when a noncommutative modification of the mass term is performed; this change involves the replacement of the classical point like behavior by a smeared distribution. In (3+1)-dimensions, the $P-v$ diagrams are quite similar to the commutative case and the critical points are found by solving a nonhomogeneous cubic equation, contrary to the classical case where the nonhomogeneous term is absent. Since this equation is more involved that in the commutative case, an explicit expression for the critical values is hard to obtain, however it can be plotted showing the existence of critical points where classically they are forbidden. The Gibbs function when plotted as a function of the temperature for fixed pressure shows the existence of critical points as well and the commutative case is also reproduced; phase transitions may or not exist depending on the values of the noncommutative parameter.

We also considered a noncommutative deformation of the source giving rise to the BI electromagnetic field. As in the first model, the relevant expressions for the mass and temperature were obtained for arbitrary dimensions and later specialized for (3+1)-dimensions. The $P-v$ diagrams have some similarities with the classical case but there are also important differences; they show a more smooth behavior. This is also seen in the plots of the function defining the critical points of the model, where noncommutativity makes this function to take higher values rapidly when compared with the classical case. In this case the Gibbs function shows a behavior similar to the Schwarzschild solution for $\theta$ and $b$ varying, nevertheless some differences can be appreciated.

\begin{figure}[pbth]
\centering
\includegraphics[width=0.45\textwidth]{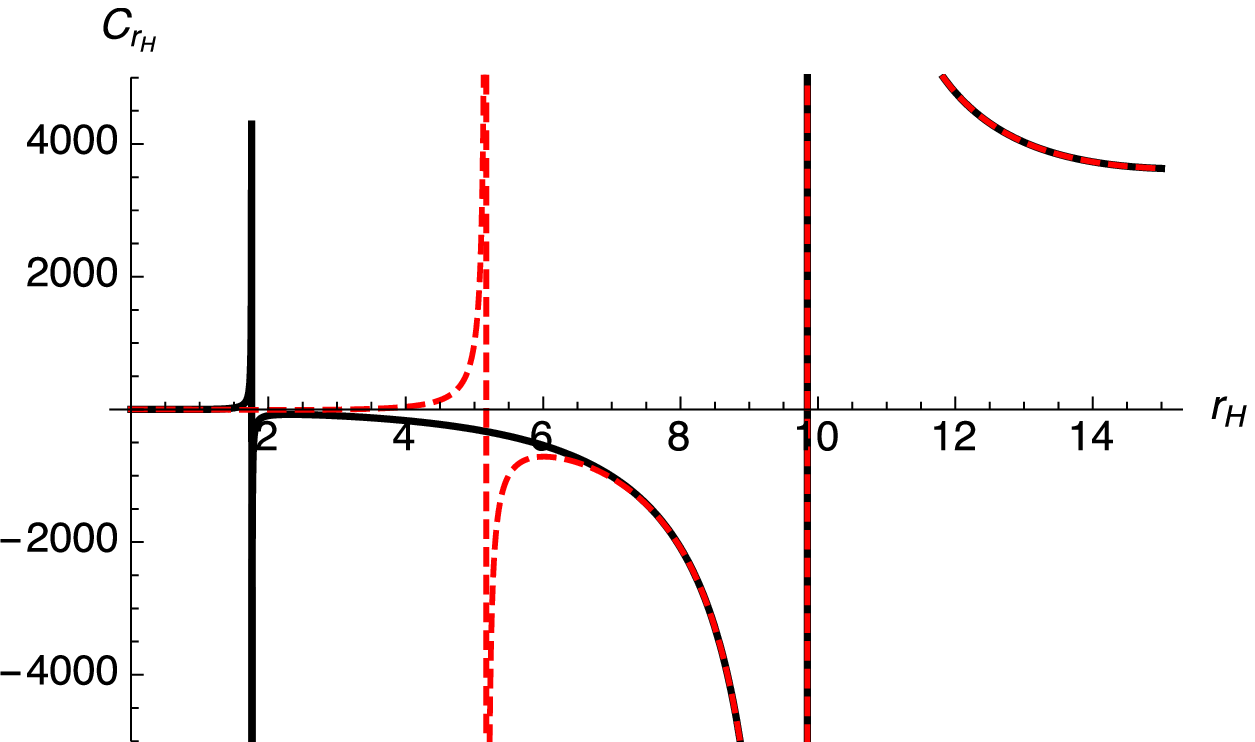}
\hfill
\includegraphics[width=0.45\textwidth]{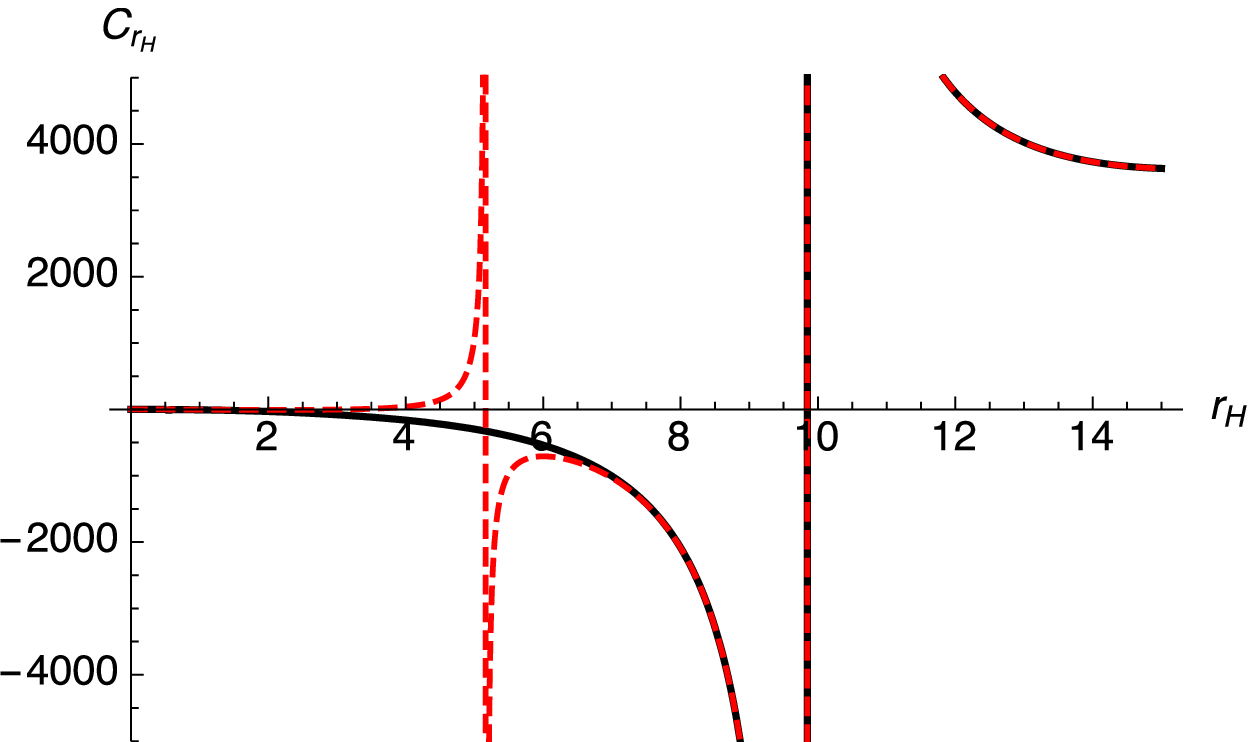}
\caption{Specific heat $C_{r_H}$ as a function of the horizon radius for $b = 100$ (left figure) and $b = 0.1$ (right figure). The solid line (black) is the curve for $\theta = 10^{-2}$ and the dashed line (red) the curve for $\theta = 1$. The values $\Lambda = - 0.01, Q = 1$ are fixed in all cases.}  
\label{fig7}
\end{figure}

From the two models analyzed in this work, model A has the interesting feature of enhancing the presence of critical points. Plots of the specific heat $C_{R_H}$ are given in Fig.~\ref{fig7} using dimensionless variables defined by the replacements $r_H \to r_S r_H, Q \to r_S Q, b \to b/r_S, \Lambda \to \Lambda/ r_S^2, \theta \to r_S^2 \theta$; here $r_S$ is a constant with dimensions of length that may be identified with the Schwarzschild radius of the source. 

\begin{figure}[pbth]
\centering
\includegraphics[width=0.45\textwidth]{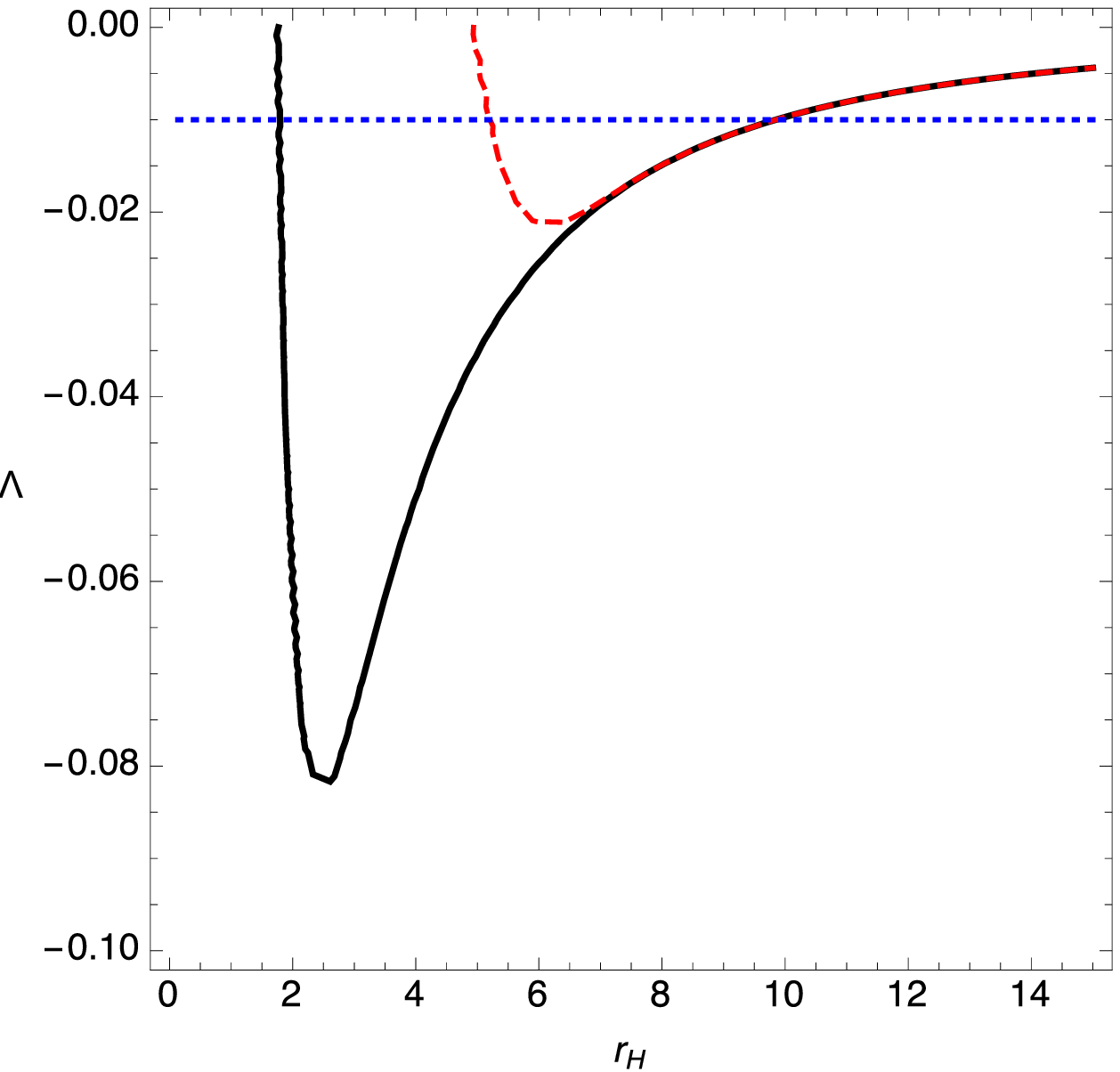}
\caption{Curves on the $r_H-\Lambda$ plane where the specific heat $C_{r_H}$ diverges for $\theta = 10^{-2}$ (black solid line) and $\theta = 1$ (red dashed line); the dotted horizontal line corresponds to the value $\Lambda = -0.01$ and the other parameters have the values $b = 100, Q = 1$.}  
\label{fig8}
\end{figure}

These graphs show the transition from large to small values of $b$ for both commutative and noncommutative cases. We notice first that for $b = 100$ the first critical point in the noncommutative case is shifted to the right compared with the standard case. Furthermore, for $b = 0.1$, the first critical point disappears if $\theta$ is small (commutative limit) while it remains for values of $\theta$ of order one. Fig.~\ref{fig8} shows the situation in the $r_H-\Lambda$ plane for $b = 100$. As $b$ takes small values, the minimum of the solid line ($\theta$ small) goes to (0, -$\infty$) and there is only one point where a horizontal line crosses this curve; the noncommutative case (dashed line) on the contrary remains essentially unchanged when going from large to small values of $b$ and have instead two intersection points with the same horizontal line. The use of the specific heat to study phase transitions in gravitational models has been done previously in the literature~\cite{Nicolini:2011dp,Banerjee:2011cz,Banerjee:2012zm,Gunasekaran:2012dq}.

The thermodynamics of a RN black hole has been studied previously~\cite{Kubiznak:2012wp} and it has been shown that this black hole behaves as a van the Wals fluid, however the effects of quantum fluctuations of spacetime are not considered. A straightforward calculation shows that in our model A for example, this result can be recovered in the limit $b\rightarrow\infty$ with ${r^2\over 4\theta } << 1$. The equation of state gets then a correcting term $M/(4\pi\theta)^{ 3/2 }$, leading to the noncommutative expression
$$
\rho_c = \frac 38 + \frac {6\pi Q^2}{(4\pi \theta)^{3/2}}.
$$
for the universal ratio of the critical exponents of the van der Wals fluid.

The techniques used in this work may in principle be applied to other kind of metrics describing for example rotating black holes with electric and magnetic charge given by a nonlinear electrodynamics~\cite{CiriloLombardo:2006ph}. It will be of interest to pursue this topic in a future work.

\acknowledgments

This research was supported by CONACyT-DFG Collaboration Grant 147492 ``Noncommutative Models in Physics". R. L. acknowledges partial support from CONACyT Grant 237351 ”Implicaciones físicas de la estructura del espacio-tiempo”. O. S.-S. was also supported by a Posdoctoral Fellowship Grant PROMEP/103.5/13/9043.

\appendix
\section{Calculation of the temperature}
\label{appa}

Here we derive a expression for the temperature of the black hole in $(3+1)$-dimensions used in the main text; its generalisation to higher dimensions is straightforward. The metric is assumed to be 
\be
ds^2 = - N^2(r) dt^2 + f^{-2}(r) + r^2 (d\theta^2 + \sin^2 \theta d\phi^2),
\ee
with $N(r) = f(r)$. Then the temperature at $r = R$ is~\cite{Brown:1994gs}
\begin{eqnarray}
T(R) &=& \left( \frac {\p E}{\p S} \right) 
\nonumber \\[4pt]
&=& \frac {\p}{\p(\pi r_H^2)} \int_{S^2} d^2 x \sqrt{\sigma} \left[ \frac 1{8\pi} \left( - \frac {2f(R)}R \right) - \epsilon_0 (R) \right]
\nonumber \\[4pt]
&=& -\frac {4\pi R^2}{8\pi R} \frac 1{f(R)} \frac {\p f^2(R)}{\p(\pi r_H^2)} = -\frac R{4\pi} \frac 1{N(R)} \frac {\p N^2(R)}{\p r_H}. 
\end{eqnarray}
Let us assume that 
\be
N^2(r) = 1 - M W(r) + V(r, Q, \Lambda),
\ee
where $W$ and $V$ are arbitrary functions of their arguments. Then the condition $N^2(r_H) = 0$ implies
\be
M = \frac 1{W(r_H)} [1 + V(r_H, Q, \Lambda)].
\ee
In consequence
\be
T(R) =  \frac 1{N(R)} \frac {R \,W(R)}{4\pi r_H} \frac {\p}{\p r_H} \frac 1{W(r_H)} [1 + V(r_H, Q, \Lambda)]. 
\ee
On the other hand, we have
\begin{eqnarray}
\frac {\p N^2(r)}{\p r} &=& -M W_{,r} + V_{,r} 
\nonumber \\[4pt]
&=& - \frac 1{W(r_H)} [1 + V(r_H, Q, \Lambda)] W_{,r} + V_{,r},
\end{eqnarray}
and evaluation at $r = r_H$ gives
\be
\frac {\p N^2(r)}{\p r} \Big |_{r_H} = W(r_H) \frac {\p}{\p r_H} \frac 1{W(r_H)} [1 + V(r_H, Q, \Lambda)].
\ee
It follows that
\begin{eqnarray}
T(R) &=& \frac 1{N(R)} \frac {R \,W(R)}{r_H W(r_H)}  \frac 1{4\pi} \frac {\p N^2(r)}{\p r} \Big |_{r_H}
\nonumber \\[4pt]
&=&  \frac 1{N(R)} \frac {R \,W(R)}{r_H W(r_H)}  \frac {\kappa_H}{2\pi}.
\end{eqnarray}
This expression was used to obtain Eq.~(\ref{temp1modela}) in the main text.


\end{document}